\documentclass [twoside,11pt]{article}
\usepackage [latin2] {inputenc}
\usepackage{graphicx}
\usepackage[normalem]{ulem} 
\usepackage{anysize}
\usepackage{fancyhdr}
\usepackage{calc}
\usepackage{geometry}
\usepackage{epsfig}
\usepackage{amssymb}
\usepackage{amsmath}
\usepackage{indentfirst}
\usepackage{chngcntr} 
\counterwithin{figure}{section}
\counterwithin{equation}{section}
\counterwithin{table}{section}
\marginsize{3,0cm}{3,0cm}{2,0cm}{2,0cm} 
\begin{document}



\begin{titlepage} 

\begin{center}
\Large  {
INSTITUTE OF PHYSICS\\
FACULTY OF PHYSICS, ASTRONOMY\\ 
AND APPLIED COMPUTER SCIENCE\\
JAGIELLONIAN UNIVERSITY }
\end{center}	

\vspace{0.5cm}

\begin{figure}[h]
\begin{center}
\includegraphics[width=3.5cm,height=4.5cm]{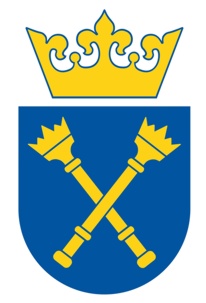}
\end{center}
\end{figure}

\vspace{0.2cm}

\begin{center}
\large{Master Thesis}
\end{center}

\vspace{0.2cm}

\begin{center}
	\Huge \textbf{Feasibility study of $\eta$-mesic \\ nuclei production\\ by means of the WASA-at-COSY \\and COSY-TOF facilities}
\end{center}
 
\vspace{0.3cm}

\begin{center}
\huge \textbf{Magdalena Skurzok}
\end{center}

\vspace{1.5cm}

\begin{center}
\LARGE{Supervisor: Prof. Pawe\l~Moskal}

\vspace{1.0cm}

\Large {Cracow, 2010}

\end{center}

\end{titlepage}


\newpage
\thispagestyle{empty}
\begin{center}
\end{center}


\newpage
\thispagestyle{empty} 
\begin{center}
\LARGE \textbf{Acknowledgments}
\end{center}

\vspace{1.0cm}

\noindent
\Large \textit{I would like to express my highest gratitude to all the people without whom this thesis would not have been possible.\\}

\Large \textit{First of all I thank Prof. Pawe\l~Moskal for valuable help, for plenty of hints and suggestions, for understanding and enormous patience. His vast knowledge and skill was the best assistance in writing this thesis.\\}

\Large \textit{I am very grateful to Prof.~Bogus\l aw Kamys for allowing me preparing this thesis in the Nuclear Physics Department of the Jagiellonian University.\\}

\Large \textit{I would like to express my appreciation to all the \mbox{WASA-at-COSY} members for their help and friendly atmosphere.\\}

\Large \textit{I also thank my colleagues:~Izabela Balwierz, Tomasz Bednarski, Szymon Nied\'zwiecki, mgr Dagmara Rozpedzik, mgr~Micha\l~Silarski, Tomasz Twar\'og and Grzegorz Wyszy\'nski for the scientific contribution and the great time spend together.\\}

\Large \textit{I would like to express my gratitude to my friends: Marcela Batkiewicz, Daria Huczek and Emilia Walczak  who were near me during all the years of my studies and encouraged me.\\}

\Large \textit{The last, but not least, I want to thank my parents and rest of my Family for the love, patience and incredible support they provided to me through my entire life.\\}


\newpage
\thispagestyle{empty}
\begin{center}
\end{center}



\newpage
\thispagestyle{empty}

\begin{center}
\Large \textbf{Abstract}

\vspace{1.0cm}

\Large \textbf{Feasibility study of~$\eta$-mesic nuclei production by means of~WASA-at-COSY and COSY-TOF facilities}
\end{center}

\vspace{1.0cm}

\large
Despite the fact that existence of $\eta$-mesic nuclei in which the $\eta$ meson might be bound with the light nucleus by means of the strong interaction was postulated already in 1986, it is still not experimentally confirmed. Discovering of this new kind of an exotic nuclear matter is very important as it might allow for better understanding of $\eta$ meson structure and its interaction with nucleons. 

\noindent 
The search of the $\eta$-helium bound states is carried out at the COSY accelerator in the Research Center J\"ulich in Germany, by means of the WASA detection system.~The search are conducted with high statistic and high acceptance for the free production of the $^{4}\hspace{-0.03cm}\mbox{He}$-$\eta$ bound states. It is also considered to search for $\eta$-tritium in quasi free reaction which might be realised with \mbox{COSY-TOF} facility.

\noindent
In this thesis the results of the Monte Carlo simulations of the $\eta$-helium bound states and $\eta$-tritium bound state are presented and discussed.~The acceptances of the \mbox{WASA-at-COSY} and \mbox{COSY-TOF} detectors for the free and quasi-free \mbox{$\eta$-mesic} nuclei production reactions were determined, respectively.~Furthermore acceptances were compared for three different models of nucleon momentum distribution inside atomic nuclei and three different values of width of a considered bound states. In case of COSY-TOF detector it was established that the most effcient measurement of quasi-free $dd\rightarrow p_{sp}(\mbox{T}$-$\eta)_{bs}\rightarrow p_{sp} d p \pi{}^{-}$ reaction can be done at beam momentum of $p_{beam}$=3.1GeV/c.   


\newpage
\thispagestyle{empty}
\begin{center}
\end{center}


\newpage
\def\contentsname{\LARGE Contents \vspace{0.5cm}}
\tableofcontents 

\newpage
\thispagestyle{empty}
\begin{center}
\end{center}



\newpage
\pagestyle{fancy}
\fancyhf{} 
\fancyhead[LE,RO]{\textbf{\thepage}}
\fancyhead[RE]{\small\textbf{{Introduction}}} 

\newpage
\thispagestyle{plain}

\section{Introduction}
\vspace{0.5cm}

\large

\begin{figure}[h!]
\begin{center}
\includegraphics[width=6.0cm,height=6.0cm]{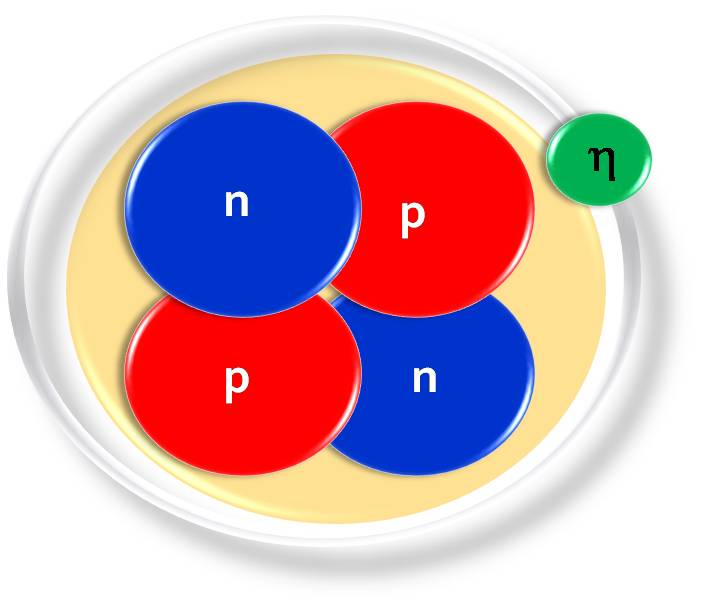}
\caption{The scheme of $\eta$-mesic bound state.}\label{Jadro}
\end{center}
\end{figure}

\noindent
The new kind of nuclear matter consisting of nucleus bound with $\eta$ mesons via strong interaction was postulated by Haider and Liu over twenty years ago~\cite{HaiderLiu1}.~However, till now none of experiments confirmed empirically its \mbox{existence}. This exotic form of matter called $\eta$-mesic nucleus is schematically presented in Fig.~\ref{Jadro}.

\indent
Up to this time $\eta$-mesic nuclei have been searched via production of $\eta$ meson in~the vicinity of the heavy nuclei.~It was considered that due to a huge number of nucleons, attraction of $\eta$ meson could be high enough that allows to form a bound state.~Nevertheless those experiments have not brought expected effect.~Recent investigations indicate that interaction between $\eta$ meson and nucleus is considerably stronger than it was predicted earlier. Therefore, according to the current theoretical considerations, it is possible that the bound states might be also formed for a light nuclei like helium, tritium~\cite{Wilkin1,WycechGreen} and even deuteron~\cite{Green}.\\
\indent
The existence of $\eta$-mesic nuclei allows to investigate interaction of the $\eta$ meson and the nucleons inside a nuclear matter.~Moreover it would provide information about $\mbox{N}^{*}(1535)$ resonance~\cite{Jido} and about $\eta$ meson properties in nuclear matter~\cite{InoueOset}, as well as about contribution of the flavour singlet component of the quark-gluon wave function of $\eta$ meson~\cite{BassThomas, BassTom}.\\  

\indent
The measurement of the $^{4}\hspace{-0.03cm}\mbox{He}$-$\eta$ bound states is carried out with a unique accurance by means of WASA detector~\cite{Adam1} installed at cooler synchrotron COSY in the Research Center J\"ulich. The $\eta$-mesic nuclei is searched there via studying of excitation function for the chosen decay channels of the \mbox{$^{4}\hspace{-0.03cm}\mbox{He}$-$\eta$} system formed in deuteron-deuteron collision~\cite{Moskal4}.~The measurement is performed for the beam momentum varying continously around the threshold.~The~beam ramping technique allows to reduce the systematical uncertainities.~The \mbox{existence} of the bound system should manifest itself as a resonance-like structure in the excitation curve of eg. $dd\rightarrow(^{4}\hspace{-0.03cm}\mbox{He}$-$\eta)_{bs}\rightarrow$ $^{3}\hspace{-0.03cm}\mbox{He} p \pi{}^{-}$ reaction below the $dd\rightarrow$ $^{4}\hspace{-0.03cm}\mbox{He}$-$\eta$ reaction threshold which allows to determine the~\mbox{binding} energy and the width of such state.
  
\indent
Formation of the $\eta$-mesic nucleus might be also realized by means of the quasi-free reactions. In this case the scan of the energy can be achieved from the Fermi motion of nucleons inside the deuteron beam. Measurements of such reaction is available for the external COSY-TOF detector~\cite{Pizzolotto,Jaeckle,Abdel} where the search of $\eta$-mesic Tritium can be carried out by the measurement of the excitation function of the  $nd\rightarrow(\mbox{T}$-$\eta)_{bs}\rightarrow$ $d p \pi{}^{-}$ reaction around the threshold of the $nd \rightarrow \mbox{T}$-$\eta$  production~\cite{Moskal4}.\\  
\indent
The main aim of this thesis is a determination of geometrical acceptance of WASA detector for four reactions in which $\eta$-mesic bound states might be formed via free production:\\

\noindent
$dd\rightarrow(^{4}\hspace{-0.03cm}\mbox{He}$-$\eta)_{bs}\rightarrow$ $^{3}\hspace{-0.03cm}\mbox{He} p \pi{}^{-}$\\
$dd\rightarrow(^{4}\hspace{-0.03cm}\mbox{He}$-$\eta)_{bs}\rightarrow d p p \pi{}^{-}$\\
$pd\rightarrow(^{3}\hspace{-0.03cm}\mbox{He}$-$\eta)_{bs}\rightarrow d p \pi{}^{0} \rightarrow d p \gamma \gamma$\\ 
$pd\rightarrow(^{3}\hspace{-0.03cm}\mbox{He}$-$\eta)_{bs}\rightarrow p p p \pi{}^{-}$\\

\noindent
and geometrical acceptance of COSY-TOF detector setup for one quasi-free reaction of $\eta$-mesic nuclei production:\\

\noindent
$dd\rightarrow p_{sp}(\mbox{T}$-$\eta)_{bs}\rightarrow p_{sp} d p \pi{}^{-}$\\

\noindent
In each case Monte Carlo simulations of $\eta$-mesic nucleus production and decay process were carried out based on reaction kinematics. The simulations were realized with assumption that the bound state has a resonance structure given by the Breit-Wigner distribution with fixed binding energy $\mbox{B}_{s}$ and a width $\Gamma$. Moreover, for reconstruction of events the spectator model was applied. The efficiency of the registration of each reactions was analysed with regard to different models describing Fermi momentum distributions of nucleons inside deuteron, helium and tritium nuclei. It was also compared for different values of $\mbox{B}_{s}$ and $\Gamma$ from range predicted by theory~\cite{GarNiIno}.

\indent       
This thesis is divided into seven chapters. The second describes indirect and direct experimental indications for the existence of the $\eta$-mesic helium.\\ 

The Chapter 3 treats of nucleon momentum distributions inside the light nuclei such as $^{3}\hspace{-0.03cm}\mbox{He}$, $^{4}\hspace{-0.03cm}\mbox{He}$ and T. The distributions are presented and compared for different models.\\

Description of the spectator model assumptions and its experimental confirmations are presented in Chapter 4.\\           
        
The Chapter 5 is devoted to the kinematics of the free and quasi-free reactions in which $\eta$-mesic bound states are produced and decay.\\

The WASA-at-COSY and COSY-TOF detection systems emphasising their properties useful for the measurement of the $\eta$-mesic nuclei are presented in Chapter 6.\\

Simulation results of the bound states in free and quasi-free reactions are described in Chapter 7 and in Chapter 8 a summary and conclusions are presented.


\pagestyle{fancy}
\fancyhf{} 
\fancyhead[LE,RO]{\textbf{\thepage}}
\fancyhead[RE]{\small\textbf{{Indications for the existence of eta-mesic helium}}} 

\newpage
\thispagestyle{plain}

\large

\newpage
\section{Indications for the existence of $\eta$-mesic helium}

\vspace{0.5cm}

\subsection{Indirect}

\large

\noindent
According to the theoretical considerations, the formation of the $\eta$-mesic nucleus can only take place if the real part of the $\eta$-nucleus scattering length is negative (attractive nature of the interaction), and the magnitude of the real part is greater than the magnitude of the imaginary part~\cite{HaiderLiu2}:

\begin{equation}
|Re(a_{\eta-nucleus})|>|Im(a_{\eta-nucleus})|.\label{eq:jeden}
\end{equation}

\noindent
A wide range of possible values of the $\eta$N scattering lenght $a_{\eta N}$ calculated for hadronic- and photoproduction of the $\eta$ meson has not exluded the formation of $\eta$-nucleus bound states for a light nuclei as $^{3,4}\hspace{-0.03cm}\mbox{He}$, T~\cite{Wilkin1,WycechGreen} and even for deuteron~\cite{Green}.~Those bound states have been searched in many experiments. However, none of them gave empirical confirmation of their existence. There are only a signal which might be interpreted as an~\mbox{indications} of the $\eta$-mesic nuclei.
         
\indent
Experimental observations which might suggest the possibility of~the existence of the $^{3}\hspace{-0.03cm}\mbox{He}$-$\eta$ bound system were found by \mbox{SPES-4}~\cite{Berger}, \mbox{SPES-2}~\cite{Mayer}, ANKE~\cite{Mersmann} and \mbox{COSY-11}~\cite{Smyrski1} collaborations. In the experiments cross section of the \mbox{$dp\rightarrow$ $^{3}\hspace{-0.03cm}\mbox{He}$-$\eta$} \mbox{reaction} was measured. In this reaction with the real $\eta$ meson in the final state, the bound state could not be produced and therefore obtained results might be treated as indirect indications of the $\eta$-mesic nuclei only. 

\begin{figure}[h!]
\centering
\includegraphics[width=8.0cm,height=7.0cm]{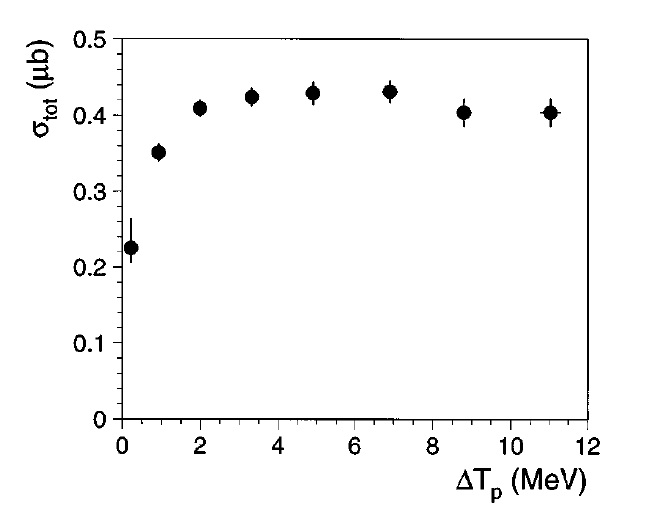}
\caption{Total cross section for $pd\rightarrow$ $^{3}\hspace{-0.03cm}\mbox{He}$-$\eta$ reaction measured for eight different proton energies above threshold from 0.2 to 11~MeV~\cite{Mayer}. Total cross section rises from 0.25 to 0.40~$\mu$b rapidly in the range of 2~MeV.\label{spes_2}} 
\label{ratio}
\end{figure}

\newpage
\indent
The~\mbox{measurements} on the SPES-4 spectrometer were realised \mbox{using} the deuteron beam accelerated in SATURNE synchrotron colliding with \mbox{liquid-hydrogen} target, while in case of SPES-2 the proton beam and \mbox{liquid-deuterium} target were used.~The total cross section measured by the SPES-2 collaboration for the eight different beam \mbox{energies} are presented in~Fig.~\ref{spes_2}.
According to \cite{Wilkin1}, the energy dependence of the cross section for the~$dp\rightarrow$ $^{3}\hspace{-0.03cm}\mbox{He}$-$\eta$ reaction is dominated by the strong interaction between $\eta$ and $^{3}\hspace{-0.03cm}\mbox{He}$ \mbox{originating} from the strong $\eta$-nucleon interaction leading to the formation of~the $\mbox{N}^*$(1535) resonance. 
The data analysis of the close to threshold measurements of the total cross \mbox{section} led to the determination of the~$\eta^{3}\hspace{-0.03cm}\mbox{He}$ scattering length. The negative sign of a real part of the $a_{\eta^{3}\hspace{-0.03cm}{He}}$ and its large value equal \mbox{$a_{\eta^{3}\hspace{-0.03cm}{He}}=(-2.31+2.57i)$}~fm~\cite{Wilkin1} suggest a possible existence of the \mbox{$(^{3}\hspace{-0.03cm}\mbox{He}$-$\eta)_{bs}$}, \mbox{although} the condition given by (\ref{eq:jeden}) is not \mbox{fulfilled}. 

\begin{figure}[h]
\centering
\includegraphics[width=6.7cm,height=6.3cm]{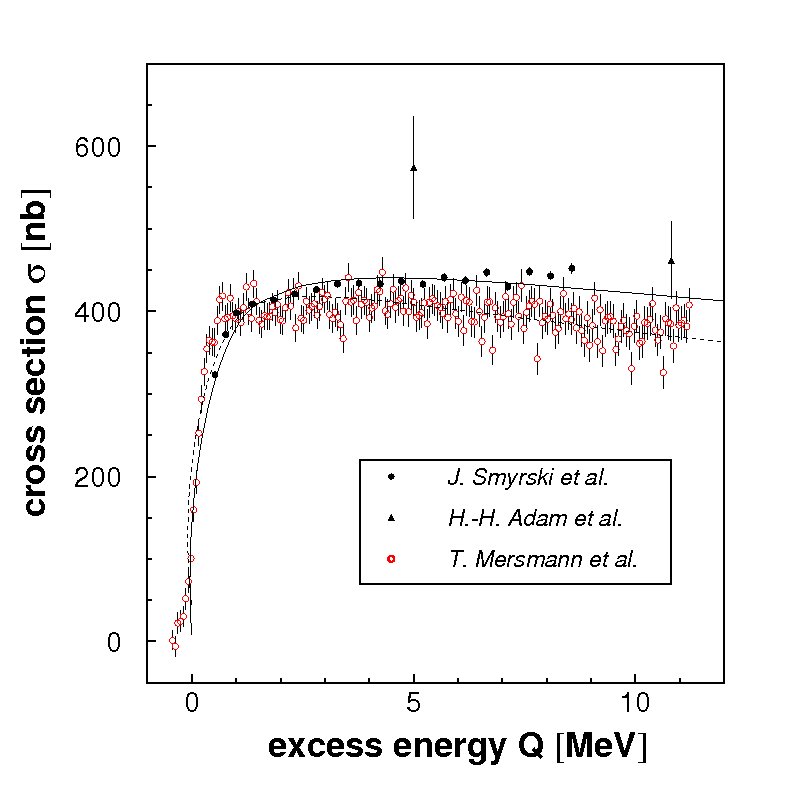}  \includegraphics[width=6.7cm,height=6.4cm]{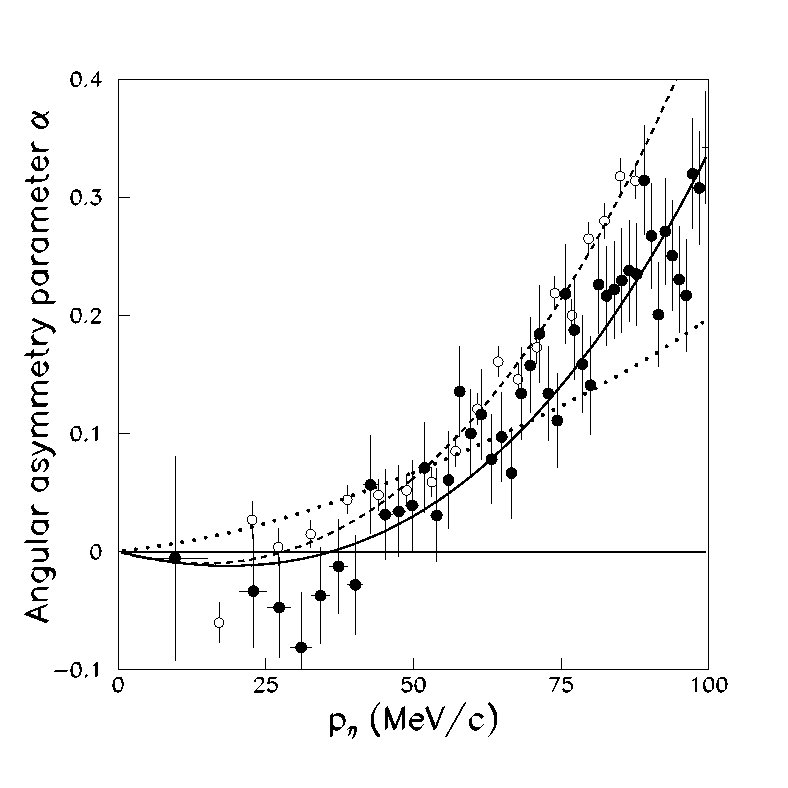}
\caption{(left) Total cross section for the $dp\rightarrow$ $^{3}\hspace{-0.03cm}\mbox{He}$-$\eta$ reaction measured with the \mbox{COSY-ANKE} (open circles)~\cite{Mersmann} and the \mbox{COSY-11} facilities (closed circles)~\cite{Smyrski1} and (triangles)~\cite{Adam}. Scattering length fit to the COSY-ANKE and COSY-11 data is represented with dashed and solid lines, respectively. (right) Angular asymmetry parameter $\alpha$ for the experimental data from COSY-ANKE (full dots)~\cite{Mersmann} and from COSY-11 (open circles)~\cite{Smyrski1}. The dashed and solid lines are fitted (assuming the phase variation between S and P waves) to the COSY-11 and COSY-ANKE data, respectively~\cite{Wilkin2}. The fit without the phase variation is denoted as the dotted line. The figure is adapted from~\cite{Moskal4}.\label{cosy11_anke}}
\end{figure}

\indent
An indirect signature of the~$\eta$-mesic nuclei were also searched at the cooler synchrotron COSY in~J\"ulich by~means~of~internal \mbox{COSY-11} and \mbox{COSY-ANKE} detection setups with high precision and high statistics. In the experiments the momentum ramping technique of the deuteron beam was used that allows to reduce the systematic uncertainties. The beam was accelerated slowly and linearly in~time, from excess energy of~\mbox{Q=-5.05 MeV} up~to \mbox{Q=11.33 MeV} in~case~of~ANKE experiment~\cite{Mersmann}, while during the \mbox{COSY-11} experiment~\cite{Smyrski1} the~momentum was varied in the range corresponding to the excess energy from \mbox{Q=-10 MeV} to \mbox{Q=9 MeV}. Both collaborations performed the measurement of~the~excitation function and differential cross section of~the \mbox{$dp\rightarrow$ $^{3}\hspace{-0.03cm}\mbox{He}$-$\eta$} reaction close to the~kinematical threshold. The experimental excitation function parametrized with the s-wave formula of scattering length~\cite{Mersmann,Smyrski1} is presented in~Fig.~\ref{cosy11_anke} (left panel). The fit to~the~\mbox{COSY-11} data gave the value of the $\eta^{3}\hspace{-0.03cm}\mbox{He}$ scattering length equal to $a_{\eta^{3}\hspace{-0.03cm}{He}}=[\pm(2.9\pm0.6)+(3.2\pm0.4)i]$~fm~\cite{Smyrski1}. Although this value is in agreement with formula (\ref{eq:jeden}), uncertainties of its real and imaginary part are too large to confirm the possible formation of \mbox{($^{3}\hspace{-0.03cm}\mbox{He}$-$\eta)_{bs}$}.~The real part of scattering lenght of the \mbox{$\eta^{3}\hspace{-0.03cm}\mbox{He}$} system derived by fitting the ANKE data for Q$<$4MeV equals $Re(a_{\eta ^{3}\hspace{-0.03cm}{He}})=(11.6\pm1.4)$~fm while imaginary part is equal $Im(a_{\eta ^{3}\hspace{-0.03cm}{He}})=(-4.1\pm7.0)$~fm~\cite{Mersmann}. Those large values implies the existence of a quasi-bound states very close to the reaction threshold, however the derived by ANKE and COSY-11 collaborations real parts of the scattering length are not consistent within the quoted errors.  

\noindent
The differential cross section measurement allows to calculate angular asymmetry parameter $\alpha$, defined as: 

\begin{equation}
\alpha=\frac{d}{d\cos\theta_{\eta}}\ln\frac{d\sigma}{d\Omega}.\label{eq:alpha}
\end{equation}

\noindent
The momentum dependence of parameter $\alpha$ can be described correctly with assumption that the phase is varying between S and P waves~\cite{Wilkin2}. In another case the discrepancy between the experimental data and theoretical description are significant. 
The momentum dependence of $\alpha$ parameter is presented in Fig.~\ref{cosy11_anke} (right). 

\vspace{0.5cm}

\subsection{Direct}

\large

\begin{figure}[h]
\centering
\includegraphics[width=14.0cm,height=4.5cm]{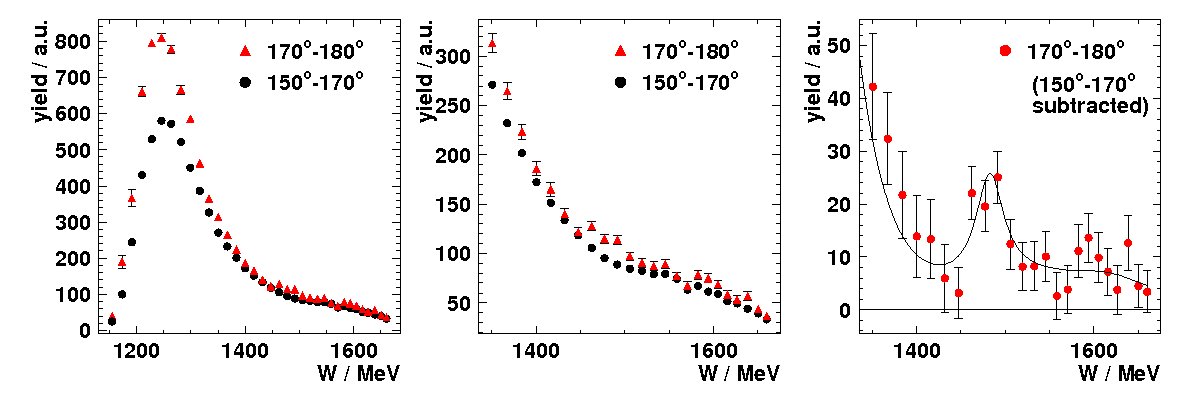}
\caption{Excitation functions of the $\pi^{0}$-proton production for relative angles of $170^{0}-180^{0}$ (red triangles) and $150^{0}-170^{0}$ (black circles) in the $\gamma^{3}\hspace{-0.03cm}\mbox{He}$ center-of-mass sytem are shown in the left and center panels. In the right panel the difference between both distributions with superimposed line denoting the results of the fit of the \mbox{Breit-Wigner} distribution plus background are presented. The figure is adapted from~\cite{Pfeiffer}.\label{mami}}
\end{figure}

\noindent
The first direct experimental indications of a light $\eta$-nucleus bound states were observed in the reaction of the $\eta$ photoproduction $\gamma^{3}$\hspace{-0.03cm}$\mbox{He}\rightarrow \pi^{0}pX$ which was investigated with the TAPS calorimeter at the electron accelerator facility Mainz Microtron (MAMI)~\cite{Pfeiffer}.~Photons produced in a~thin radiator foil and tagged using a special spectrometer hit the target filled with liquid $^{3}\hspace{-0.03cm}\mbox{He}$. There the~measurements of~the excitation functions of~the~$\pi^{0}$-proton production for two ranges of~the~relative angle between those particles were carried~out.~It appeared that a~difference between excitation curves for opening angles of~$170^{0}-180^{0}$ and $150^{0}-170^{0}$ in~the~center-of-mass frame revealed an~enhancement just below the threshold of the $\gamma^{3}\hspace{-0.03cm}\mbox{He}\rightarrow$ $^{3}\hspace{-0.03cm}\mbox{He}$-$\eta$ reaction which was interpreted as a~possible signature of a~{$^{3}\hspace{-0.03cm}\mbox{He}$-$\eta$} bound state where $\eta$ meson captured by one of~nucleons inside helium forms an intermediate $S_{11}(1535)$ resonance which decays into $\pi^{0}$-$p$ pair.~A~binding energy and width for~the anticipated quasibound $\eta$-mesic state in~$^{3}\hspace{-0.03cm}\mbox{He}$ were deduced from the~fit~of~the~\mbox{Breit-Wigner} distribution function~\cite{Pfeiffer} to the~experimental points and equal $(-4.4\pm4.2)$~MeV and $(25.6\pm6.1)$~MeV, respectively. Those values are consistent with expectations for $\eta$-mesic nuclei.~Above cited excitation functions and the difference of~them with a \mbox{Breit-Wigner} distribution and background fitted to~the~data are shown in~Fig.~\ref{mami}.\\\\

\noindent
However, due to the low statistics of the measurement the results might be interpreted not as an indication of the bound state but rather as a virtual state what is in details described in Ref.~\cite{Hanhart}. The interpretation is still under discussion~\cite{Sibirtsev}. Moreover at the recent meeting it was shown that the result may be an artefact due to the strong influence of the resonances on the shape of the  excitation function~\cite{Krusche}.\\

\begin{figure}[h]
\centering
\includegraphics[width=4.5cm,height=4.0cm]{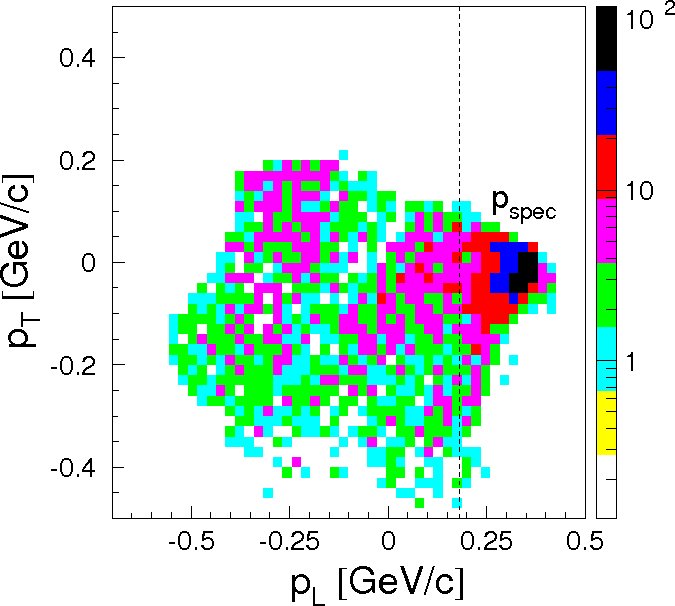} \includegraphics[width=4.1cm,height=4.0cm]{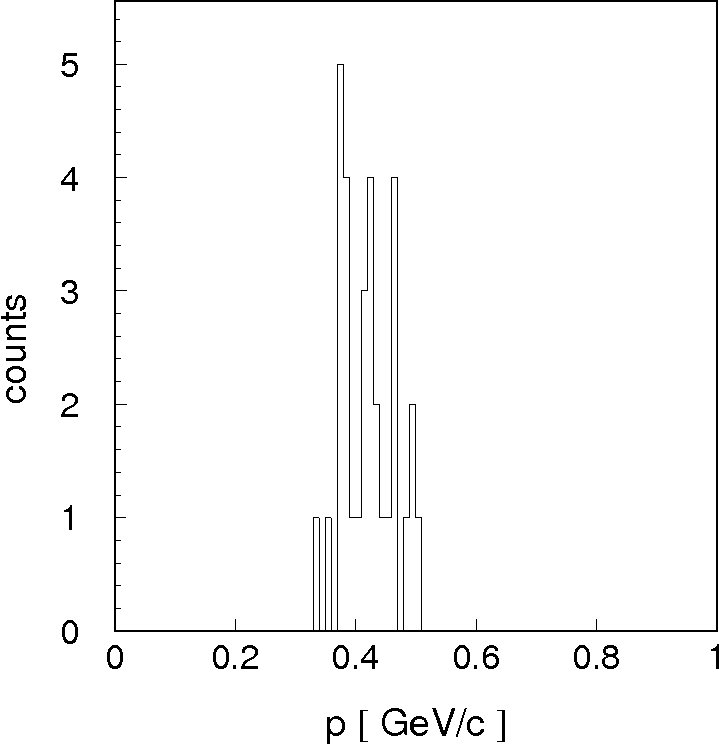} \includegraphics[width=4.1cm,height=4.0cm]{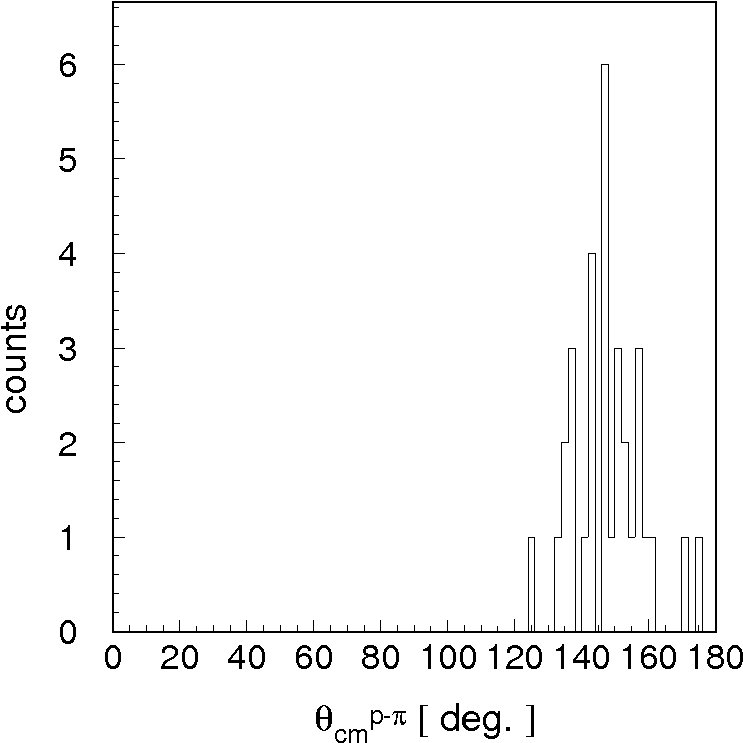}
\caption{Experimental results of the COSY-11 collaboration for the $dp\rightarrow ppp\pi^{-}$ reaction: (left) Transversal vs. longitudinal momentum distributions of protons. The upper limit for the longituidal proton momenta is shown as dashed line and equals $p_{L}$=0.18GeV/c. (middle) Pion momentum distribution in the center of mass system. (right) Relative angle between pion and proton direction in the c.m. The figure is adapted from~\cite{Krzemien1}. \label{cosy11}}
\end{figure}   

\indent
The analysis carried out by COSY-11 group~\cite{SmyMosKrze1,Krzemien1,Smyrski2,Smyrski3} give an indication for the~\mbox{$^{3}\hspace{-0.03cm}\mbox{He}$-$\eta$} bound state existence.~The search for the $\eta$-mesic helium was carried out using a deuteron beam and internal hydrogen target. The beam momentum was ramped around the kinematical threshold for the $\eta$ production in~the~$dp\rightarrow$ $^{3}\hspace{-0.03cm}\mbox{He}$-$\eta$ reaction and the measurement of the $dp\rightarrow ppp\pi^{-}$ and \mbox{$dp\rightarrow$ $^{3}\hspace{-0.03cm}\mbox{He} \pi^{0}$} reactions was carried out.~In the first case the momentum distribution of the $\pi^{-}$ (middle panel in Fig.~\ref{cosy11}) and the relative angle distribution between pion and proton momentum vectors (right panel in Fig.~\ref{cosy11}) were determined after application of apropriate cuts on the momentum of the spectator protons (left panel in Fig.~\ref{cosy11}) and the rejection of the events corresponding to quasi-free $\pi^{-}$ production. 
Obtained results are in agreement with theoretical expectations for particles originating from decay of the $\mbox{N}^{*}$(1535) resonance which is created as a result of the absorption of the bound $\eta$ meson in the neutron inside \mbox{$^{3}\hspace{-0.03cm}\mbox{He}$}. Based on the above results the upper limit of total cross section for the~\mbox{$dp\rightarrow(^{3}\hspace{-0.03cm}\mbox{He}$-$\eta)_{bs}\rightarrow p p p\pi^{-}$} reaction was estimated to the value of 270~nb. Similarly, investigation of the \mbox{$dp\rightarrow$ $^{3}
\hspace{-0.03cm}\mbox{He}$-$\pi^{0}$} reaction give only the value of the total upper limit of cross section of the~$dp\rightarrow(^{3}\hspace{-0.03cm}\mbox{He}$-$\eta)_{bs}\rightarrow$ $^{3}\hspace{-0.03cm}\mbox{He} \pi^{0}$ reaction equal to 70~nb.\\


\newpage
\thispagestyle{plain}

\section{Nucleon momentum distributions inside d, T, $^{3}\mbox{He}$ and $^{4}\mbox{He}$ nuclei}

\vspace{0.5cm}
\pagestyle{fancy}
\fancyhf{} 
\fancyhead[LE,RO]{\textbf{\thepage}}
\fancyhead[RE]{\small\textbf{{Nucleon momentum distributions inside d, T, $^{3}\mbox{He}$ and $^{4}\mbox{He}$ nuclei}}} 

\noindent
Due to the Fermi motion, nucleons inside atomic nuclei are not at rest but move with momenta which vary in a broad range.
This variation influences kinematics of nuclear reactions.
 
\begin{figure}[h]
\centering
\includegraphics[width=11.0cm,height=10.0cm]{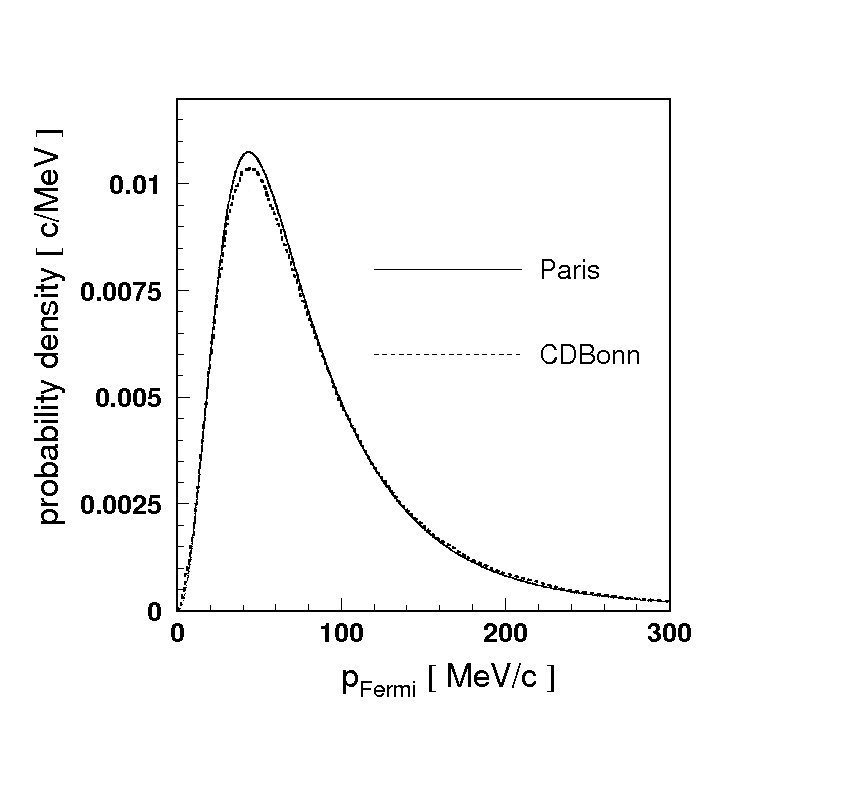}
\caption{Fermi momentum distribution of nucleons inside the deuteron for PARIS (full line) and CD-Bonn (dashed line) potentials. The distributions were normalized to unity in the momentum range from 0 to 300 MeV/c.\label{deuteron}}
\end{figure}

\indent
Fermi momentum distributions of proton and neutron bound inside a deuteron derived from two different potential models, namely PARIS~\cite{Lacombe} and CD-Bonn~\cite{Machleidt} are shown in Fig.~\ref{deuteron}.~The normalized nucleon momentum distributions are calculated by means of the Fourier transformation of parametrized deuteron wave functions obtained from space representation.~Respective parametrization coefficients found in the analitic representation of the deuteron wave function for both of above named nucleon-nucleon interaction models are given in~\cite{Lacombe,Machleidt,Czyzykiewicz}. 
The momentum distributions deduced from Paris and \mbox{CD-Bonn} potentials are peaked at about 40 MeV/c and differ no more than $5\%$~\cite{Czyzykiewicz}. 

\begin{figure}[h]
\includegraphics[width=9.0cm,height=8.5cm]{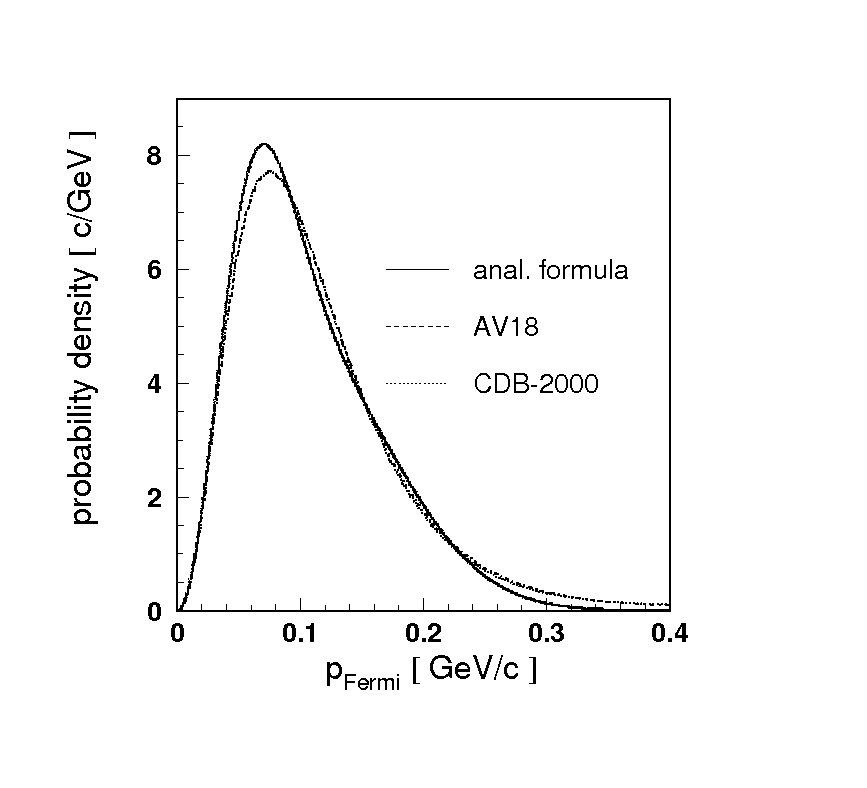} \hspace{-1.5cm}\includegraphics[width=9.0cm,height=8.5cm]{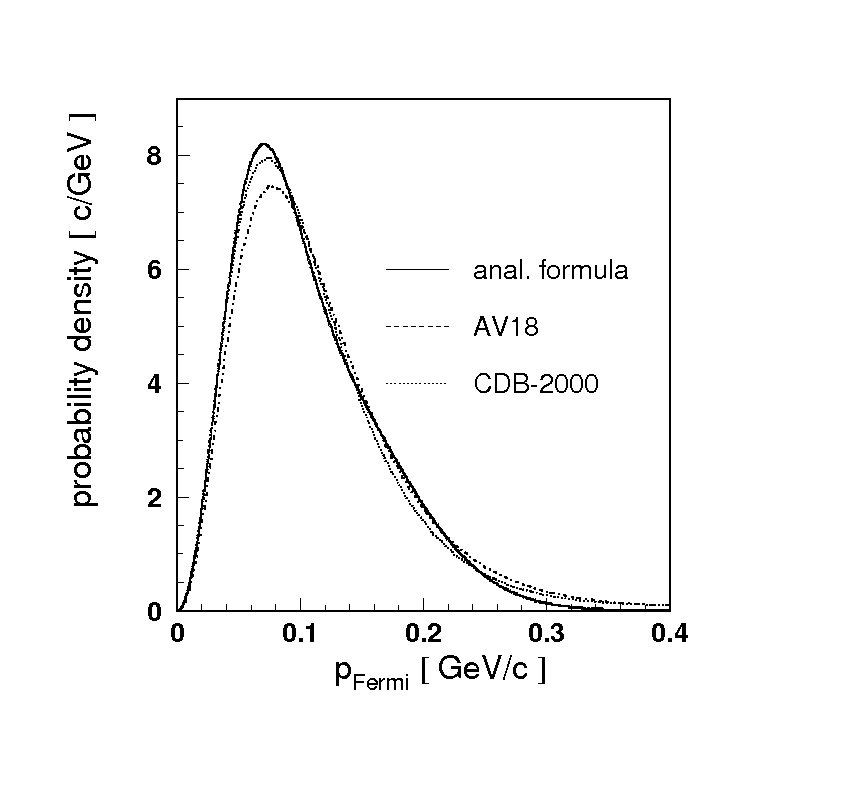}
\caption{Fermi momentum distribution for protons inside $^{3}\hspace{-0.03cm}\mbox{He}$ (left) and neutrons inside T (right) given by analytic formula (thick line) and estimated for the AV18 NN (dashed line) and the CDB-2000 NN (dotted line).~The distributions were normalized to unity in the momentum range from 0 to 0.4 GeV/c.\label{hel_3}}
\end{figure}

In case of three-nucleon bound states like $^{3}\hspace{-0.03cm}\mbox{He}$ and T, Fermi momentum distributions of nucleons are presented in Fig.~\ref{hel_3} for three different models. Thick solid line depicts proton momentum distribution inside $^{3}\hspace{-0.03cm}\mbox{He}$ and neutron momentum distribution inside T as given by analytic formula (\ref{eq:4.1}) which results from the fit to the experimental data on $p\left(^{3}\hspace{-0.03cm}\mbox{He},2p\right)d$ and $p\left(\mbox{T},pn\right)d$ reactions~\cite{Abdullin}:

\begin{equation}
f{(p)}=p^2[exp(-263p^2)+0.177exp(-69.2p^2)] 
\label{eq:4.1}
\end{equation}

\noindent
This momentum distribution is in a good agreement with the one calculated with realistic potential in the frame of the model of the composite quark bags~\cite{Yu}.
 
The distributions of protons and neutrons momentum inside $^{3}\hspace{-0.03cm}\mbox{He}$ and T are also estimated based on the AV18 and the CDB-2000 nucleon-nucleon interaction models in conjunction with Urbana IX (UIX) and Tucson-Melbourne (TM) three nucleon interactions (TNI), respectively~\cite{Nogga1,Nogga2}.~They are presented in Fig.~\ref{hel_3} for protons inside $^{3}\hspace{-0.03cm}\mbox{He}$ (left) and neutrons inside T (right) and ticked as a dashed and dotted lines. Similar results can be obtained for neutron inside helium and proton inside tritium. The difference between those two distributions are small and results from different interaction Hamiltonian forms defined for above-cited models.

\noindent
Estimation based upon the formula (\ref{eq:4.1}) is consistent with the one derived from AV18 NN and CDB-2000 NN models with an accurancy better than 9\% for protons inside $^{3}\hspace{-0.03cm}\mbox{He}$ and 11\% for neutrons inside T. 

\begin{figure}[h]
\centering
\includegraphics[width=11.0cm,height=10.0cm]{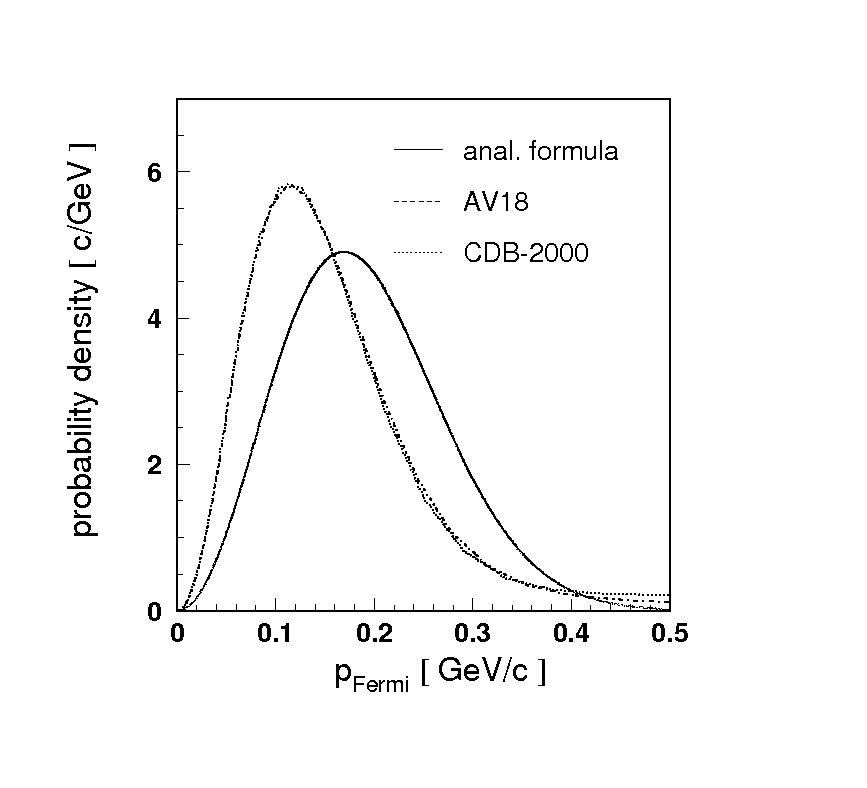}
\caption{Fermi momentum distribution of nucleons inside $^{4}\hspace{-0.03cm}\mbox{He}$ given by analytic formula (thick solid) and estimated for the AV18 NN (dashed) and the CDB-2000 NN (dotted). The distributions were normalized to unity in the momentum range from 0 to 0.5 GeV/c.~\label{hel_4}}
\end{figure}

\indent 
For nucleons inside $^{4}\hspace{-0.03cm}\mbox{He}$ Fermi momentum distributions predicted by three independent models are shown in Fig.~\ref{hel_4}. The distribution represented by a thick line is calculated from helium wave function derived based on Fermi three parameter charge distribution of nucleus~\cite{Hejny}. The momentum distribution is described by formula (\ref{eq:4.2}): 

\begin{equation}
f{(p)}=\frac{p^2}{a\cdot b}exp\left(\frac{-p^2}{a\cdot c}\right), 
\label{eq:4.2}
\end{equation}

\noindent
where $a=0.03892719$, $b=0.05511$, $c=0.7352$. Fermi momentum is given in units of GeV/c.

\indent
The dashed and dotted lines depict distributions obtained, similarly as in case of three nucleon systems, from AV18 and the CDB-2000 potential models with the inclusion of three nucleon interaction contributions~\cite{Nogga2}. Due to the fact that $^{4}\hspace{-0.03cm}\mbox{He}$ is symmetrical, proton and neutron momentum distributions are in good approximation equal.

\indent
The difference between the distributions derived from AV18 and \mbox{CDB-2000} models and given by analytic formula is significant and equals up to about 40\%, and in addition the maxima of these distributions are shifted \mbox{by about 45 MeV/c.}
The discrepancy results from the fact that the formula (\ref{eq:4.2}) was derived from nucleus charge distribution smeared out by the charge distribution of protons, whereas the AV18 and the CDB-2000 models allow for the finite size of nucleus charge distributions and are related to the momentum of the point like protons in the alpha particle~\cite{Nogga3}. Therefore, as more realistic in further considerations the momentum distributions of nucleons inside $^{4}\hspace{-0.03cm}\mbox{He}$ calculated from AV18 and the CDB-2000 potentials will be taken into account.\\

\indent
The numerical data describing above cited momentum distributions are given in \mbox{Appendix~A}. 

\indent
The presented distributions estimated based on various models will be used to calculate the systematical uncertainty in the determination of the excitation curves used to search for $\eta$-mesic bound states. Respective results are presented in Chapter~7.


\newpage
\thispagestyle{plain}
\section{Spectator model}

\vspace{0.5cm}

\subsection{The main assumption of the spectator model}

\pagestyle{fancy}
\fancyhf{} 
\fancyhead[LE,RO]{\textbf{\thepage}}
\fancyhead[RE]{\small\textbf{{Spectator model}}} 

\noindent
The production and decay of $\eta$-mesic bound states investigated in this thesis might be schematically depicted as:\\

\hspace{0.3cm}
a) $dd\rightarrow$ $(^{4}\hspace{-0.03cm}\mbox{He}$-$\eta)_{bs} \rightarrow$ 
\vspace{-0.8cm}
\begin{displaymath}
\hspace{-1.0cm}
\left\{ \begin{array}{ll}
^{3}\hspace{-0.03cm}\mbox{He}_{sp}\ p\ \pi{}^{-} \\
\ p_{sp}\ d_{sp}\ p\ \pi{}^{-}
\end{array} \right.
\end{displaymath}

\vspace{0.5cm}  
\hspace{0.3cm}   
b) $pd\rightarrow$ $(^{3}\hspace{-0.03cm}\mbox{He}$-$\eta)_{bs} \rightarrow$
\vspace{-0.8cm}
\begin{displaymath}
\hspace{-1.1cm}
\left\{ \begin{array}{ll}
p_{sp}\ p_{sp}\ p\ \pi{}^{-} \\
d_{sp}\ p\ \pi{}^{0}
\end{array} \right.
\end{displaymath}

\vspace{0.5cm}  
\hspace{0.3cm}  
c) $dd\rightarrow$ $p_{sp} n d \rightarrow$ $p_{sp} (\mbox{T}$-$\eta)_{bs}\rightarrow$ $p_{sp}\ d_{sp}\ p\ \pi{}^{-}$ \\

\vspace{0.5cm}  

\noindent  
The a) and b) schemes include free and c) one describes quasi-free production of the bound states. Subscript \textit{bs} denotes 'bound state' whereas \textit{sp} stands for the 'spectator'.\\
\indent
'Spectators' are particles which do not take part in reactions~\cite{JKlaja,Moskal1} but hit the detectors with the Fermi momentum transformed into laboratory system. Fermi momentum distributions of nucleons inside the light nuclei are presented in previous chapter.~In the framework of the spectator model due to the relatively small binding energy of the nuclei, spectators are considered as a real particles registered in the experiments and in the analysis it is assumed that they are on their mass-shell during the reaction~\cite{JKlaja,Moskal1}: 
 
\begin{equation}
\left|\mathbb{P}_{sp}\right|^2 = m_{sp}^2.
\label{eq4:4.2_1}
\end{equation}   

\noindent 
The $\mathbb{P}_{sp}$ and $m_{sp}$ are the four-momentum vector of spectator and the spectator mass, respectively.

\begin{figure}[h]
\centering
\includegraphics[width=12.0cm,height=5.0cm]{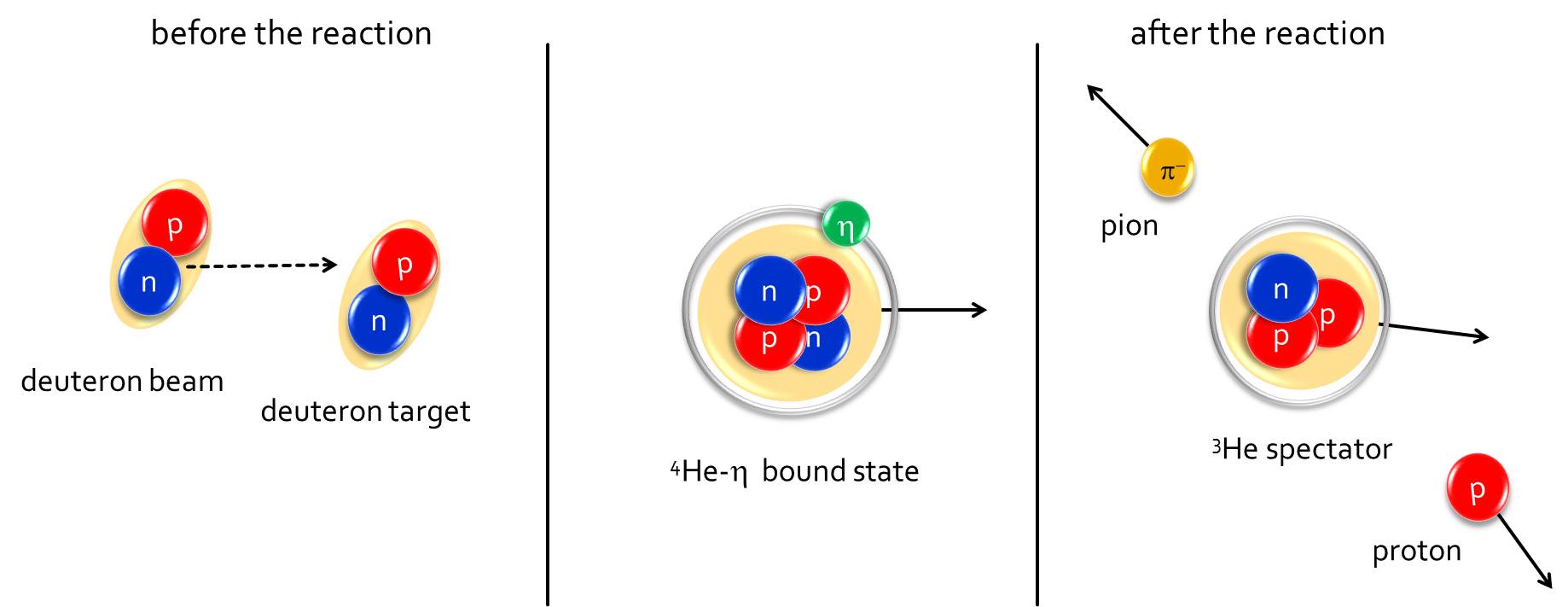} 
\caption{Schematic picture of the $dd\rightarrow(^{4}\hspace{-0.03cm}\mbox{He}$-$\eta)_{bs}\rightarrow$ $^{3}\hspace{-0.03cm}\mbox{He} p \pi{}^{-}$ reaction. Red and blue circles represent protons and neutrons respectively, whereas $\pi^{-}$ meson is depicted as yellow circle. The beam momentum is presented by the dashed arrow.\label{free}}
\end{figure}

In the free reactions the beam and target nuclei collide and form $\eta$-mesic bound state which decays into proton, pion and spectator/spectators which energies and momenta are measured in experiment. One of those reactions is schematically shown in Fig.~\ref{free}.  

\begin{figure}[h]
\centering
\includegraphics[width=12.0cm,height=5.0cm]{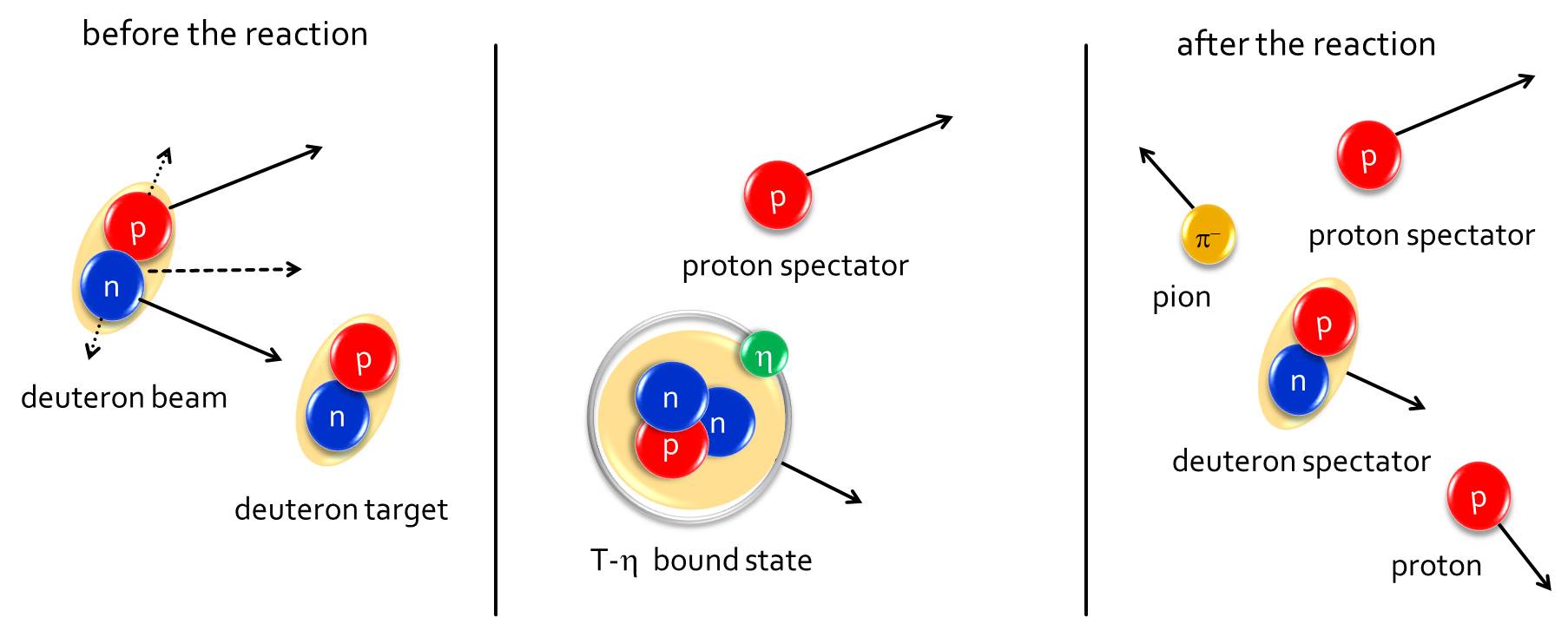} 
\caption{Schematic picture of the quasi-free $dd\rightarrow p_{sp}(\mbox{T}$-$\eta)_{bs}\rightarrow p_{sp} d p \pi{}^{-}$ reaction. Red and blue circles represent protons and neutrons respectively, whereas $\eta$ meson is depicted as green circle. The Fermi momentum of the nucleons inside the deuteron is  presented by the dotted arrows and the beam momentum by the dashed one.\label{quasi}}
\label{fig:1}
\end{figure}

In case of quasi-free reaction presented in Fig.~\ref{quasi}, the deuteron from the beam is considered as a system consisting of proton and neutron moving with the Fermi motions. For the deuteron beam we have:

\begin{equation}
\mathbb{P}_{d}=\mathbb{P}_{n}^{b} + \mathbb{P}_{p_{sp}}
\label{eq:4.2_2}
\end{equation}    

\begin{equation}
\left|\mathbb{P}_{p_{sp}}+\mathbb{P}_{n}^{b}\right|^2 = m_{d}^2,
\label{eq:4.2_22}
\end{equation}    

\newpage

\noindent
where $m_{d}$ denotes the deuteron mass equal to $1875.6$ $\mbox{MeV/c}^{2}$~\cite{Groom} while $\mathbb{P}_{p_{sp}}$ and $\mathbb{P}_{n}^{b}$ are the four-momentum vectors of the proton spectator and the beam neutron, respectively.

\noindent
According to the spectator model, proton from the beam does not take part in the reaction and is registered as a real particle whereas neutron being off the mass-shell hits the target deuteron. From the conservations of momentum and energy and the assumption that proton is on its mass-shell we obtain in the deuteron beam rest frame:
 
\begin{equation}
\vec{p_{n}^{*}}=-\vec{p_{sp}^{*}},
\label{eq:4.2_3}
\end{equation}

\begin{equation}
E_{n}^{*}=m_{d}-E_{p_{sp}}^{*},
\label{eq:4.2_4}
\end{equation}\\  

\noindent
Based on above-mentioned relationship we can deduce the neutron four-momentum vector from the spectator momentum which is fundamental in the analysis of reaction kinematics.~The close description of free and quasi-free reactions processes is described in Chapter~5. 


\vspace{0.5cm}

\subsection{Experimental proofs of spectator model}

\begin{figure}[h]
\centering
\includegraphics[width=7.2cm,height=7.0cm]{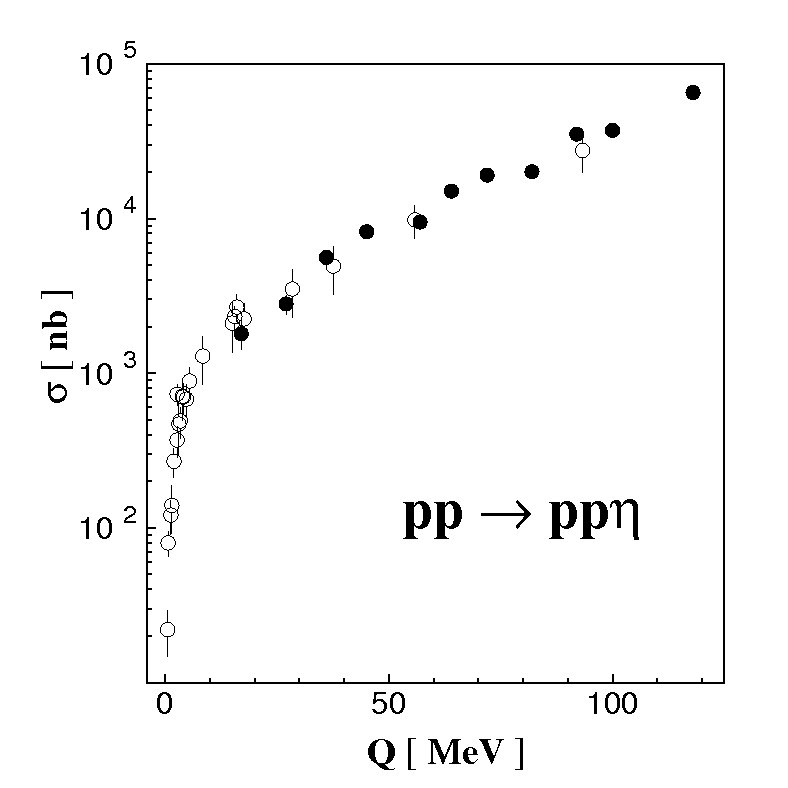} 
\caption{Total cross section for the  $pp \rightarrow$ $pp\eta$ reaction as a function of the excess energy for free (open circles) and quasi-free proton scattering (full circles). Figure is adapted from~\cite{MoskalWolke}. The data are taken from references~\cite{Smyrski4,Hibou,Calen1,Chiavassa,Bergdolt,Moskalx,Calen2}.\label{wasa_promice}}
\end{figure}

\large
\noindent
The validity of spectator model assumptions was confirmed by measurements performed by collaborations WASA/PROMICE~\cite{Stepaniak}, TRIUMF~\cite{Duncan}, \mbox{COSY-TOF}~\cite{AbdelBary}, COSY-11~\cite{Przerwa} and HADES~\cite{Hades}.~The WASA/PROMICE collaboration~\cite{Stepaniak} has compared free and quasi-free production cross sections for the $pp\rightarrow$ $pp\eta$ reaction. As a result it was presented that within the statistical errors there is no difference between the total cross section of the free and quasi-free process. The experimental data are shown in Fig.~\ref{wasa_promice}.  

Investigation of pion production at the TRIUMF~\cite{Duncan} facility in quasi-free \mbox{$pp\rightarrow$ $d\pi^{+}$} reaction extracted from the \mbox{$pd\rightarrow$ $d\pi^{+}n$} reaction has proven that the spectator momentum distribution determined from the experimental data agrees with expectations based on spectator model. Moreover, it was shown that the magnitude of the differential cross sections for the quasi-free and for the free reactions are consistent on the few per cent level. 

The spectator assumption was also confirmed by the COSY-TOF group~\cite{AbdelBary}. The momentum distribution of the spectator as well as the shape of the angular distribution for the quasi-free $np\rightarrow$ $pp\pi^{-}$ and $pn\rightarrow$ $pn$ reactions have been measured. The experimental data are consistent with calculations based upon the hypothesis of spectator model with an accurancy better than 4\% up to 150 MeV/c of the Fermi momentum and with about 25\% up to a momentum of~300 MeV/c. 

\begin{figure}[h!]
\centering
\includegraphics[width=7.5cm,height=7.0cm]{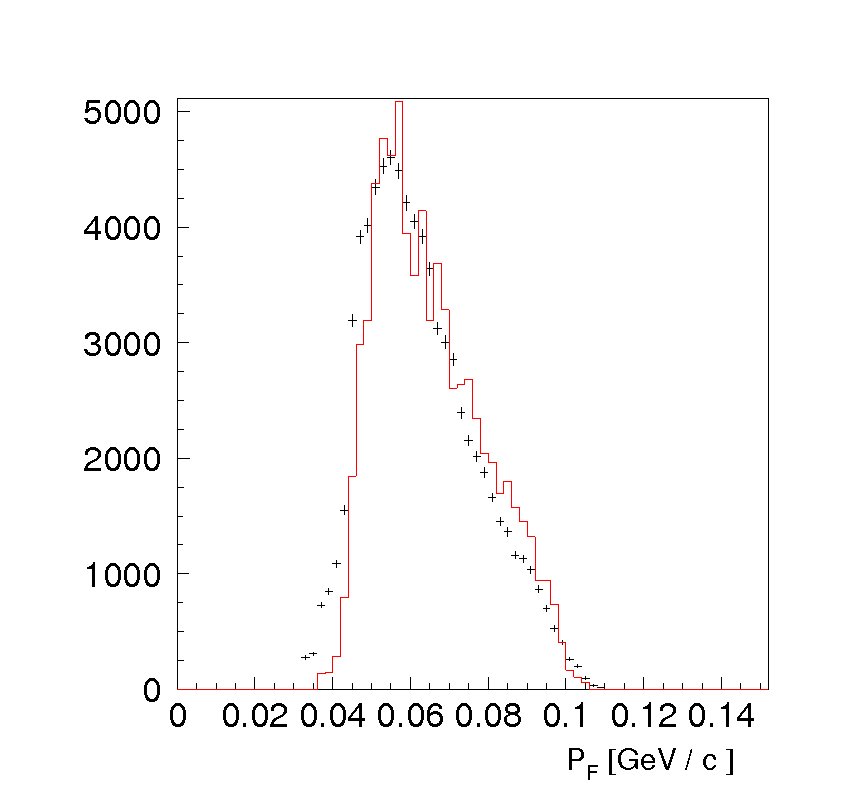} 
\caption{Proton spectator momentum distribution reconstructed in COSY-11 experiment (points) in comparison with simulation taking into account Fermi momentum distribution of nucleons inside deuteron, the acceptance and the efficiency of the detector system (solid histogram). The figure is adapted from~\cite{Przerwa}.\label{cosy-11}}
\label{ratio}
\end{figure}

In case of quasi-free $pn\rightarrow$ $pn\eta'$ reaction studied at COSY-11 facility~\cite{Przerwa} it is shown that the measured proton spectator momentum distribution is in good agreement with the theoretical assumptions of spectator model. The comparison of the experimental data and simulation result is shown in Fig.~\ref{cosy-11}. 
 
\begin{figure}[h!]
\centering
\includegraphics[width=13.0cm,height=6.5cm]{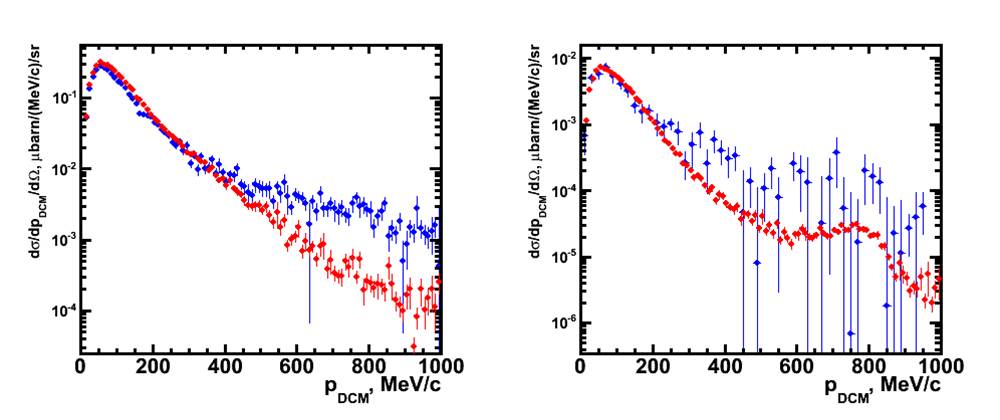} 
\caption{Momentum distribution of proton spectator measured in HADES experiment (red points) and calculated based upon the hypothesis of spectator model (blue points) for \mbox{$M_{e^{+}e^{-}}<140 \mbox{MeV/c}^{2}$} (left) and for \mbox{$M_{e^{+}e^{-}}>140 \mbox{MeV/c}^{2}$}(right). Picture courtesy of~\cite{Hades}.\label{HADES}}
\label{ratio}
\end{figure}

Recently, the validity of the spectator model was proven also by HADES collaboration~\cite{Hades} during the measurement of the quasi-free $np\rightarrow$ $e^{+} e^{-} p X$ reaction realised with a deuteron beam and proton target. The angular distribution of proton spectator outgoing from deuteron beam and its momentum distribution agrees with assumptions of spectator model up to 200-300 MeV/c. Momentum distributions in deuteron CM frame are presented in~Fig.~\ref{HADES}. Experimental data are ticked as a red points, while simulations results based on the spectator model as a black points. 

The above-cited results confirmed the spectator model and thus allow to use it in the analysis of the reactions kinematics. 

\newpage


\pagestyle{fancy}
\fancyhf{} 
\fancyhead[LE,RO]{\textbf{\thepage}}
\fancyhead[RE]{\small\textbf{{Kinematics of the $\eta$-mesic bound states production and decays}}} 

\newpage
\thispagestyle{plain}
\section{Kinematics of the $\eta$-mesic bound states production and decays}

\vspace{0.5cm}

\large
\noindent
In this thesis four reactions of free and one of quasi-free $\eta$-mesic bound states production are considered~\cite{Moskal2}:

\begin{enumerate}
\item $dd\rightarrow(^{4}\hspace{-0.03cm}\mbox{He}$-$\eta)_{bs}\rightarrow$ $^{3}\hspace{-0.03cm}\mbox{He} p \pi{}^{-}$
\item $dd\rightarrow(^{4}\hspace{-0.03cm}\mbox{He}$-$\eta)_{bs}\rightarrow d p p \pi{}^{-}$
\item $pd\rightarrow(^{3}\hspace{-0.03cm}\mbox{He}$-$\eta)_{bs}\rightarrow d p \pi{}^{0}$ $\rightarrow d p \gamma \gamma$ 
\item $pd\rightarrow(^{3}\hspace{-0.03cm}\mbox{He}$-$\eta)_{bs}\rightarrow p p p \pi{}^{-}$
\item $nd\rightarrow(\mbox{T}$-$\eta)_{bs}\rightarrow d p \pi{}^{-}$
\end{enumerate}

\begin{figure}[h!]
\centering
\includegraphics[width=11.5cm,height=6.0cm]{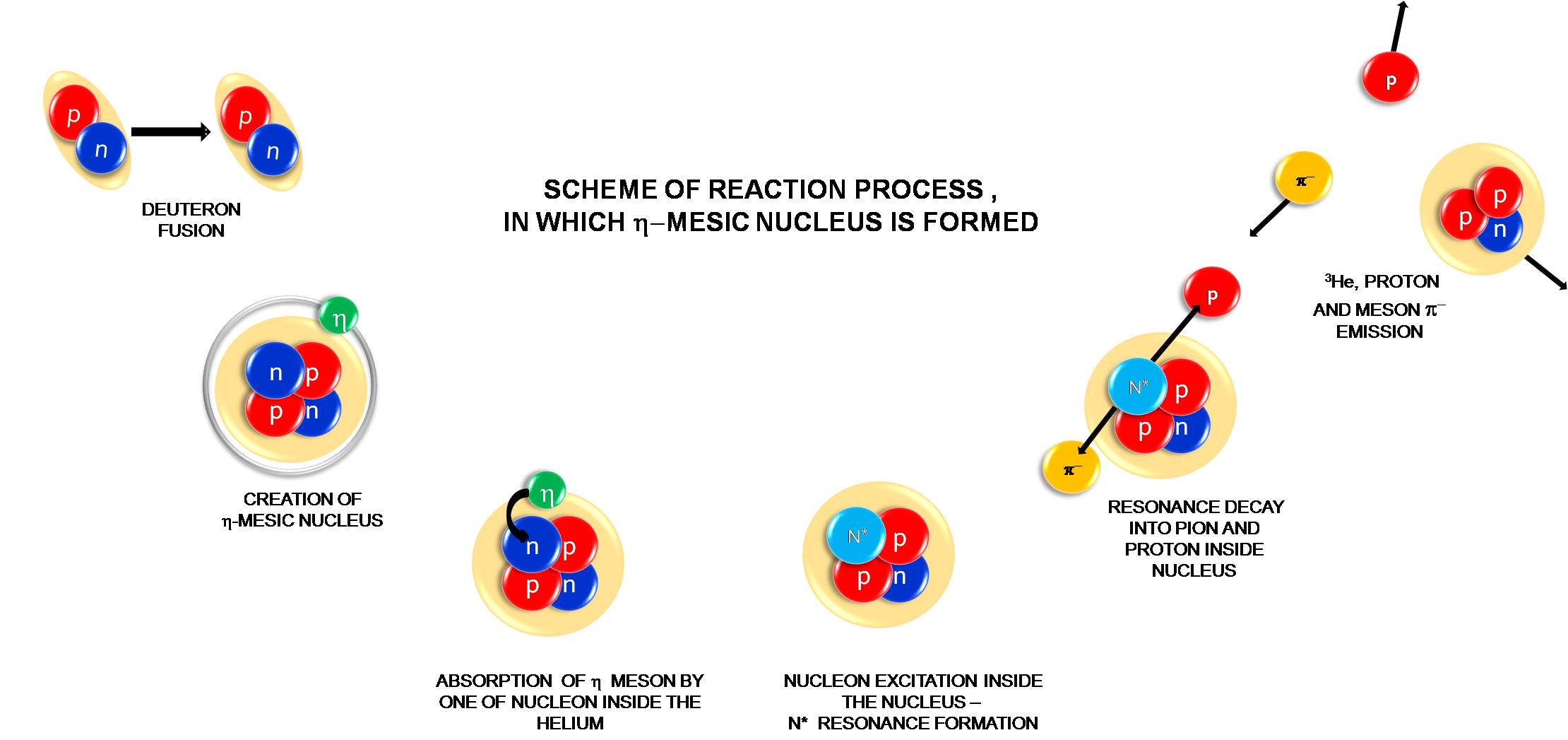}
\caption{Reaction process of the ($^{4}\hspace{-0.03cm}\mbox{He}$-$\eta)_{bs}$ production and decay.\label{free_reaction}}
\end{figure}

\noindent 
In case of free reactions (1)-(4), $^{4}\hspace{-0.03cm}\mbox{He}$-$\eta$ and $^{3}\hspace{-0.03cm}\mbox{He}$-$\eta$ bound states are produced in deuteron-deuteron and proton-deuteron fusion, respectively~\cite{Moskal3}.~The mechanism of the reactions is presented schematically in the example of the $(^{4}\hspace{-0.03cm}\mbox{He}$-$\eta)_{bs}$ production in Fig.~\ref{free_reaction}.~Describing the kinematics of the reaction following notations will be used:\\

\noindent
$\mathbb{P}_{d}^{\,b}=(E_{d}^{\,b},\vec{p}_{b})$-four-momentum vector of the beam deuteron\\
$\mathbb{P}^{\,t}_{d}=(m_{d},0)$-four-momentum vector of the target deuteron\\
$\mathbb{P}_{^{3}\hspace{-0.05cm}He}=(E_{^{3}\hspace{-0.05cm}He},\vec{p}_{^{3}\hspace{-0.05cm}He})$-four-momentum vector of the outgoing $^{3}\hspace{-0.03cm}\mbox{He}$\\
$\mathbb{P}_{p}=(E_{p},\vec{p}_{p})$-four-momentum vector of the outgoing proton\\
$\mathbb{P}_{\pi^{-}}=(E_{\pi^{-}},\vec{p}_{\pi^{-}})$-four-momentum vector of the outgoing pion

\newpage
According to the scheme shown in Fig.~\ref{free_reaction}, the deuteron from the beam hits the deuteron in the target with a momentum of $\vec{p}_{b}$.~The collision may lead to the creation of $^{4}\hspace{-0.03cm}\mbox{He}$ nucleus bound with the $\eta$ meson via strong interaction. The mass of a bound state is a sum of $\eta$ and $^{4}\hspace{-0.03cm}\mbox{He}$ masses reduced by binding energy ($\mbox{B}_{s}$):

\begin{equation}
m_{bs}=m_{\eta}+m_{^{4}\hspace{-0.05cm}He}-B_{s}.
\end{equation}

\noindent
The $\eta$-mesic nucleus moves in laboratory frame with velocity:

\begin{equation}
\vec{\beta}_{cm}=\frac{\vec{p}_{b}}{m_{d}+E_{d}^{\,b}}=\frac{2\,\vec{p}_{b}\,m_{d}}{s_{dd}},
\end{equation}

\noindent
where $s_{dd}$ is the square of invariant mass of the colliding deuterons: 

\begin{equation}
s_{dd}=|\mathbb{P}_{d}^{\,b}+\mathbb{P}_{d}^{\,t}|^{2}=2m_{d}\left(m_{d}+\sqrt{m^{2}_{d}+|\vec{p_{b}}|^{2}}\right).
\end{equation}

\noindent
The $\eta$ meson might be absorbed by one of the nucleons inside helium and may propagate in the nucleus via consecutive excitation of nucleons to the $\mbox{N}^{*}(1525)$ state~\cite{Sokol} until the resonance decays into the pion-proton pair outgoing from the nucleus~\cite{Moskal4,Moskal3,KrzeMosSmy}. Before the decay, it is assumed that $\mbox{N}^*$ resonance moves with a Fermi momentum $\vec{p}^{\,\,*}_{F}$ inside $^{4}\hspace{-0.03cm}\mbox{He}$. From the momentum conservation in the $^{4}\hspace{-0.03cm}\mbox{He}$ frame and the assumption of spectator model, momentum and energy of $^{3}\hspace{-0.03cm}\mbox{He}$ may be expressed as:

\begin{equation}
\vec{p}^{\,\,*}_{^{3}\hspace{-0.05cm}He}=-{\vec{p}}^{\,\,*}_{F}
\end{equation}     

\begin{equation}
E^{\,*}_{^{3}\hspace{-0.05cm}He}=\sqrt{m^{2}_{^{3}\hspace{-0.05cm}He}+|\vec{p}^{{\,\,*}}_{F}|^{2}}.
\end{equation}\\     

\noindent
The momentum and energy are transformed into the laboratory frame by means of Lorentz transformation:

\begin{equation}
\vec{p}_{^{3}\hspace{-0.05cm}He}=\vec{p}^{\,\,*}_{^{3}\hspace{-0.05cm}He}+\vec{\beta}_{cm}\gamma_{cm}(\gamma_{cm}/(\gamma_{cm}+1)\vec{\beta}_{cm}\cdot\vec{p}^{\,\,*}_{^{3}\hspace{-0.05cm}He}+E^{\,*}_{^{3}\hspace{-0.05cm}He})
\end{equation}

\begin{equation}
E_{^{3}\hspace{-0.05cm}He}=\gamma_{cm}(E^{\,*}_{^{3}\hspace{-0.05cm}He}+\vec{\beta}_{cm}\cdot\vec{p}^{\,\,*}_{^{3}\hspace{-0.05cm}He}),
\end{equation}\\

\noindent
where $\gamma_{cm}=1/\sqrt{1-|\vec{\beta}_{cm}|^{2}}.$\\

\noindent
The angle between outgoing $^{3}\hspace{-0.03cm}\mbox{He}$ and the beam direction is given by:

\begin{equation}
\theta_{^{3}\hspace{-0.05cm}He}=\arccos{\left(\frac{\vec{p}_{^{3}\hspace{-0.05cm}He}\cdot\vec{p}_{b}}{|\vec{p}_{^{3}\hspace{-0.05cm}He}|\cdot|\vec{p}_{b}|}\right).}
\end{equation}

\noindent
The relative angle between the outgoing nucleon-pion pair is equal to $180^\circ$ in the $\mbox{N}^{*}$ reference frame. In the following the variables in the N* reference frame will be denoted by '**'. Both particles move with a momentum $|\vec{{p}^{{\,\,**}}_{p,\pi^{-}}}|$ which is related to the resonance mass: 

\begin{equation}
m_{{N}^*}=\left(s_{dd}+m^{2}_{^{3}\hspace{-0.05cm}He}-2\sqrt{s_{dd}}\sqrt{m^{2}_{^{3}\hspace{-0.05cm}He}+|\vec{p}^{{\,\,*}}_{F}}|^{2}\right)^{\frac{1}{2}},
\label{eq:10}
\end{equation}\\

\noindent
and is given by: 

\begin{equation}
|\vec{p}^{\,\,**}_{p,\pi^{-}}|=\frac{\lambda(m^{2}_{{N}^*},m^{2}_{{\pi}^{-}},m^{2}_{{p}})}{2m_{{N}^*}},
\label{eq:101}
\end{equation}\\

\noindent
where $\lambda(x,y,z)=(x-y-z)^{2}-4yz$~\cite{Byckling}.\\

\noindent
The pion and proton four-momentum vectors in the laboratory frame are calculated using the Lorentz transformation, first from $\mbox{N}^{*}$ to the bound state frame:

\begin{equation}
\vec{p}^{\,\,*}_{p,\pi^{-}}=\vec{p}^{\,\,**}_{p,\pi^{-}}+\vec{\beta}_{N^{*}}\gamma_{N^{*}}(\gamma_{N^{*}}/(\gamma_{N^{*}}+1)\vec{\beta}_{N^{*}}\cdot\vec{p}^{\,\,**}_{p,\pi^{-}}+E^{\,**}_{p,\pi^{-}})
\label{eq:11}
\end{equation}

\begin{equation}
E^{\,*}_{p,\pi^{-}}=\gamma_{N^{*}}(E^{\,**}_{p,\pi^{-}}+\vec{\beta}_{N^{*}}\cdot\vec{p}^{\,\,**}_{p,\pi^{-}}),
\label{eq:12}
\end{equation}\\

\noindent
and further to the laboratory frame:

\begin{equation}
\vec{p}_{p,\pi^{-}}=\vec{p}^{\,\,*}_{p,\pi^{-}}+\vec{\beta}_{cm}\gamma_{cm}(\gamma_{cm}/(\gamma_{cm}+1)\vec{\beta}_{cm}\cdot\vec{p}^{\,\,*}_{p,\pi^{-}}+E^{\,*}_{p,\pi^{-}})
\label{eq:13}
\end{equation}

\begin{equation}
E_{p,\pi^{-}}=\gamma_{cm}(E^{\,*}_{p,\pi^{-}}+\vec{\beta}_{cm}\cdot\vec{p}^{\,\,*}_{p,\pi^{-}}),
\label{eq:14}
\end{equation}\\

\noindent
where $\gamma_{N^{*}}=1/\sqrt{1-|\vec{\beta}_{N^{*}}|^{2}}$ is a velocity of the resonance $\mbox{N}^{*}$ in the bound state frame.\\

\noindent
The angles of outgoing proton and pion in LAB system equals:

\begin{equation}
\theta_{p,\pi^{-}}=\arccos{\left(\frac{\vec{p}_{p,\pi^{-}}\cdot\vec{p}_{b}}{|\vec{p}_{p,\pi^{-}}|\cdot|\vec{p}_{b}|}\right)}.
\label{eq:15}
\end{equation}\\

The quasi-free reaction kinematics was partially characterized in Chapter~4 by the way of the spectator model description. 
Neutron bound inside the deuteron hits the deuteron target forming the (T-$\eta)_{bs}$, while proton does not take part in reaction and is considered as a real particle:

\begin{equation}
|\mathbb{P}_{p_{sp}}|^{2}=m^{2}_{p_{sp}}.  
\end{equation}
 
\noindent 
The square of invariant mass of neutron-deuteron system ($s_{nd}$) depends on the neutron Fermi momentum $\vec{p}^{\,\,*}_{n}$ (5.1.3) inside the beam deuteron and taking into account that target deuteron is at rest in the laboratory $s_{nd}$ equals:

\begin{equation}
s_{nd}=(E_{n}+m_{d})^{2}-|\vec{p}_{n}|^{2},  
\end{equation}

\noindent
where $\vec{p}_{n}$ and $E_{n}$ are neutron momentum and energy in laboratory frame which may be obtained applying the Lorentz transformation according to the following formulas:

\begin{equation}
\vec{p}_{n}=\vec{p}^{\,\,*}_{n}+\vec{\beta}_{d}\gamma_{d}(\gamma_{d}/(\gamma_{d}+1)\vec{\beta}_{d}\cdot\vec{p}^{\,\,*}_{n}+E^{*}_{n})
\label{eq:6.1}
\end{equation}
 
\begin{equation}
E_{n}=\gamma_{d}(E^{*}_{n}+\vec{\beta}_{d}\cdot\vec{p}^{\,\,*}_{n}),
\label{eq:6.2}
\end{equation}\\

\noindent
where $\vec{\beta}_{d}$ denotes velocity of the beam deuteron in the laboratory frame, and $\vec{p}^{\,\,*}_{n}$ denotes neutron momentum in the deuteron center of mass, and $E^{*}_{n}=m_{d}-\sqrt{m^{2}_{p}+|p^{*}_{n}|^{2}}$.\\

\noindent
The process of T-$\eta$ bound state decay is analogous like in case of the free reaction which kinematics was described before. The deuteron spectator escapes with the Fermi momentum and energy, which in the frame of the bound state is equal to:

\begin{equation}
\vec{p}^{\,\,*}_{d}=-{\vec{p}}^{\,\,*}_{F}
\end{equation}     

\begin{equation}
E^{\,*}_{d}=\sqrt{m^{2}_{d}+|\vec{p}^{{\,\,*}}_{F}|^{2}},
\end{equation}\\     

\noindent
After the transformation into laboratory system we have:

\begin{equation}
\vec{p}_{d}=\vec{p}^{\,\,*}_{d}+\vec{\beta}_{cm'}\gamma_{cm'}(\gamma_{cm'}/(\gamma_{cm'}+1)\vec{\beta}_{cm'}\cdot\vec{p}^{\,\,*}_{d}+E^{\,*}_{d})
\end{equation}

\begin{equation}
E_{d}=\gamma_{cm'}(E^{\,*}_{d}+\vec{\beta}_{cm'}\cdot\vec{p}^{\,\,*}_{d}),
\end{equation}\\

\noindent
where $\vec{\beta}_{cm'}$=$\frac{\vec{p}_{n}}{m_{d}+E_{n}}$ denotes velocity of the center of mass for the quasi-free $nd\rightarrow(\mbox{T}$-$\eta)_{bs}\rightarrow d p \pi{}^{-}$ reaction in the laboratory frame.

\indent
The outgoing proton-pion pair originates from the decay of the resonance created via absorption of the $\eta$ meson on a nucleon in the tritium nucleus. The four-momenta of those particles are described in resonace frame by equations analogous to (\ref{eq:11}) and (\ref{eq:12}), and in laboratory frame by formulas (\ref{eq:13}) and (\ref{eq:14}) while proton and pion angles with relation to the beam direction are given by (\ref{eq:15}).


\pagestyle{fancy}
\fancyhf{} 
\fancyhead[LE,RO]{\textbf{\thepage}}
\fancyhead[RE]{\small\textbf{{Experimental setup}}} 

\newpage
\thispagestyle{plain}
\section{Experimental setup}

\vspace{0.5cm}

\subsection{WASA-at-COSY facility}

\begin{figure}[h]
\centering
\includegraphics[width=15.0cm,height=8.5cm]{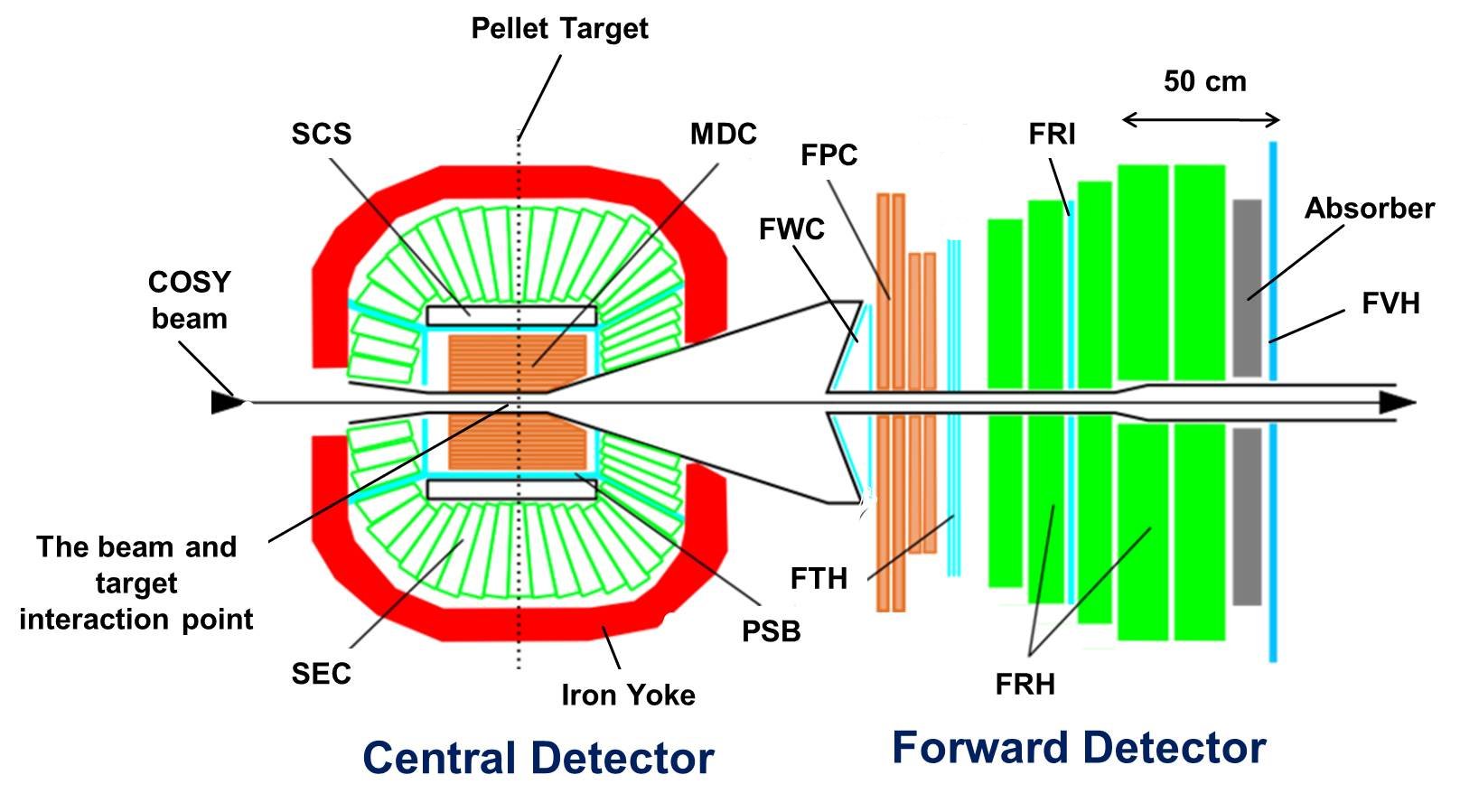}
\caption{Scheme of WASA-at-COSY detection system. Gamma quanta, electrons and charged pions being products of mesons decays are registered in the Central Detector. Scattered projectiles and charged recoil particles like $^{3}\hspace{-0.03cm}\mbox{He}$, deuterons and protons are registered in the Forward Detector. The abbreviations of the detectors names are explained in the text.\label{wasa_detector}}
\end{figure}

\large
\noindent
The search for $\eta$-mesic helium in free production reactions with high statistic and acceptance is carried out at the WASA facility~\cite{Adam1}, an internal detection system installed at the cooler synchrotron COSY in the Research Center J\"ulich. The WASA detector vertical cross section is schematically presented in Fig.~\ref{wasa_detector}.~All setup components and the method of~measurement are described in detail in reference~\cite{Adam1}. Thus, in this chapter the experimental technique will be only shortly presented.\\

In the COSY synchrotron protons and deuterons might be accelerated in~the momentum range between 0.3 GeV/c and 3.7 GeV/c~\cite{Adam1}. The ring can be filled with up to $10^{11}$ particles leading to luminosities of $10^{31} \mbox{cm}^{-2}\mbox{s}^{-1}$ in~case of internal cluster target~\cite{Brauksiepe} and $10^{32} \mbox{cm}^{-2}\mbox{s}^{-1}$ in case of pellet target~\cite{Adam1}. Beams are cooled by means of electron cooling as well as stochastic cooling at injection and high energies, respectively.

\indent
The internal hydrogen ($\mbox{H}_{2}$) or deuteron ($\mbox{D}_{2}$) target of the pellet-type is \mbox{installed} in the central part of the WASA-at-COSY detector and its position is marked in Fig.~\ref{wasa_detector} as a dotted line. The central detector is built around the interaction point and designed for measurements of $\pi^{0}$ and $\eta$ mesons decay products like \mbox{photons}, electrons and charged pions.~The charged particles momenta and reaction \mbox{vertex} are determined by means of Mini Drift Chamber (MDC) which covers angles from 24$^{0}$ to 159$^{0}$.~Charged \mbox{particles} are here bending in the magnetic field provided by \mbox{sourrounding} Superconducting Solenoid (SCS). First their trajectories are reconstructed, and then knowing the magnetic field, the momentum vector is reconstructed. For identification of charged particles the $\Delta$E-p and $\Delta$E-E methods are used based on $\Delta$E signals in Plastic \mbox{Scintillator} Barrel (PSB). The photons, electrons and positrons are \mbox{registered} in Scintillator Electromagnetic Calorimeter (SEC) via production of \mbox{electromagnetic} cascades. The calorimeter covers polar angle in the range from 20$^{0}$ to 169$^{0}$. \\  
\indent
The detection and identification of forward scattered projectiles and target-recoil particles such as protons, deuterons and He nuclei and also of neutrons and charged pions are carried out with the Forward Detector which covers the range of polar angles from 3$^{0}$ to 17$^{0}$. It consists of fourteen planes of plastic scintillators forming  Forward Window Counter (FWC), \mbox{Forward} Trigger Hodoscope (FTH), Forward Range Hodoscope (FRH), Forward Range Interleaving Hodoscope (FRI) and Forward Veto Hodoscope (FVH), respectively and proportional counter drift tubes called Forward Proportional Chamber (FPC).
Particles trajectories are reconstructed from the signals registered successively in FWC, FPC, FTH and FVH scintillator modules. Particles are identified based on measurement of energy loss in~the detection layers of FRH, FWC and FTH. The registered energy loss allows to determine their total momentum which direction is reconstucted from the measurement of~particles tracks by means of straw detectors constituting FPC. Respective components of the Forward Detector are presented in Fig.~\ref{wasa_detector}.\\
\indent    
The $^{3}\hspace{-0.03cm}\mbox{He}$-$\eta$ and $^{4}\hspace{-0.03cm}\mbox{He}$-$\eta$ bound states considered in this thesis can be searched at WASA-at-COSY detection setup in proton-deuteron and deuteron-deuteron fusion reaction, respectively. The measurement will be carried out for the beam momentum slowly ramped around the $\eta$ production threshold corresponding to the range of excess energy Q from about -60 MeV to 20~MeV. The existence of the $\eta$-mesic nucleus should be visible in the excitation function as a \mbox{resonance-like} structure below the He-$\eta$ production threshold.~The free \mbox{$\eta$-helium} bound states production reactions will be carried out in experiment based on measurement of four-momenta of the outgoing particles. WASA \mbox{detector} at COSY allows for simultaneous registration of all ejectiles with large \mbox{acceptance}, which eg. for~the detection of the $dd\rightarrow(^{4}\hspace{-0.03cm}\mbox{He}$-$\eta)_{bs}\rightarrow$ $^{3}\hspace{-0.03cm}\mbox{He} p \pi{}^{-}$ \mbox{reaction} equals about 60\%. \mbox{Spectators} from the reactions will be registered mainly in the Forward \mbox{Detector}, while \mbox{proton-pion} pair from the resonance decay will be registered for the most part in the \mbox{Central} \mbox{Detector}. The detailed description of the reactions kinematics is given in Chapter~5.
  
\vspace{0.5cm}
\subsection{COSY-TOF facility}

\begin{figure}[h]
\centering
\includegraphics[width=15.0cm,height=10.0cm]{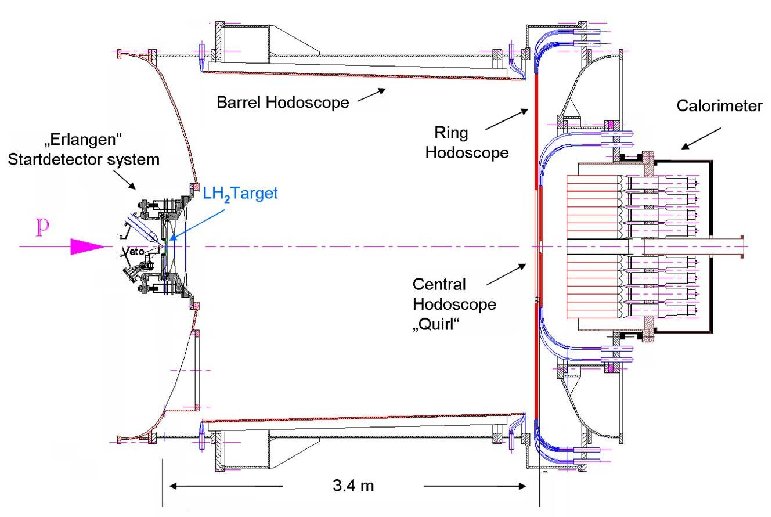}
\caption{Scheme of COSY-TOF detection setup. Charged particles trajectories are registered in 'Erlangen' start detector system, while their identification is carried out with the Barrel, Ring and Quarrel detectors. The figure is adapted from~\cite{Abdel}\label{cosy_tof}.}
\end{figure}

\noindent
The measurement of $\eta$-mesic tritium might be realised by means of the \mbox{quasi-free} reaction with the time-of-flight spectometer COSY-TOF, a~'4$\pi$~\mbox{detector'} \mbox{installed} at an external beamline of the COSY synchrotron. The detection setup is schematically presented in Fig.~\ref{cosy_tof}. Detailed description of each particular parts and the measurement technique might be found in~\cite{Pizzolotto,Abdel,AbdelBary}.

The proton or deuteron beam accelerated in COSY is extracted and hits a target which contains liquid hydrogen or liquid \mbox{deuterium}~\cite{Pizzolotto, Jaeckle}. The target is installed in front of the 'Erlangen' start detector system \mbox{consisting of} four modular detectors (Starttorte, Microstrip detector, Small Hodoscope and Large Hodoscope) designed for a precision geometric charged particles tracks reconstruction. The detector covers a polar angular range from 3.4$^{0}$ to 74$^{0}$. The time-of-flight of the charged particles outgoing after the interaction of the beam particles with the target, is measured  with an accuracy of 0.25 ns by means of the stop detector situated in cylindric vacuum tank. This detector consists of Barrel detector as well as of two Endcap detectors called Quirl and Ring and covers angular range from 0.7$^{0}$ to 76.7$^{0}$. Knowing the time-of-flight and the flight length between start and stop detectors, ~particles velocity is determined. The particle momentum can be calculated from the velocity and the mass hypothesis. Additionally, neutral particles which take part in the reactions and have not been detected can be analysed via momentum and energy conservation. 

\indent
The search of $\eta$-tritium bound state can be carried out at COSY-TOF facility via the \mbox{measurement} of the excitation function of the \mbox{$nd\rightarrow(\mbox{T}$-$\eta)_{bs}\rightarrow d p \pi{}^{-}$} reaction.~The signal from $(\mbox{T}$-$\eta)_{bs}$ is expected below the threshold of~the \mbox{$nd \rightarrow \mbox{T}$-$\eta$} production~\cite{Moskal4,Moskal2}.~In the experiment deuteron beam will be used and the $nd$ reaction will be analized based on the measurement of the four-momentum of~spectator proton ($p_{sp}$)~\cite{Moskal4} from the \mbox{$dd\rightarrow p_{sp} n d\rightarrow p_{sp}(\mbox{T}$-$\eta)_{bs}\rightarrow p_{sp} d p \pi{}^{-}$} reaction, which was schematically shown in Fig~\ref{quasi}. The advantage of this quasi-free reaction is that the Fermi momentum distribution of nucleons inside the deuteron beam allows for the scan of energy around the $\eta$ meson production threshold at a fixed value of the beam momentum.
Deuterons, protons and pions being products of T-$\eta$ bound states decays will be detected in a multi layer scintillator detectors by measuring their time of flight as well as direction.\\  


\newpage
\pagestyle{fancy}
\fancyhf{} 
\fancyhead[LE,RO]{\textbf{\thepage}}
\fancyhead[RE]{\small\textbf{{Simulation results}}} 

\thispagestyle{plain}
\section {Simulation results}

\noindent
In this chapter simulation results of the free and quasi-free $\eta$-mesic bound states production are presented. Monte-Carlo calculations were carried out by means of computer programme written in FORTRAN'90 language. 

\vspace{0.5cm}

\subsection{Simulation program scheme}
\noindent
The main purpose of simulations was the determination of the geometrical acceptances of WASA-at-COSY and \mbox{COSY-TOF} detectors for free and quasi-free reactions, respectively and comparing them for different models of nucleon momentum distribution inside atomic nuclei and different values of a bound states width.

In case of free reactions the simulation might be schematicaly described in following points for example of reaction (1) which kinematics is presented in details in Chapter~5:

\begin{enumerate}
\item The square of invariant mass of the whole system $\sqrt{s_{dd}}$ is distributed randomly according to the Breit-Wigner distribution which is given by formula~(\ref{eq:BW}) and shown in Fig.~\ref{BW}.

\begin{figure}[h]
\centering
\includegraphics[width=11.0cm,height=9.0cm]{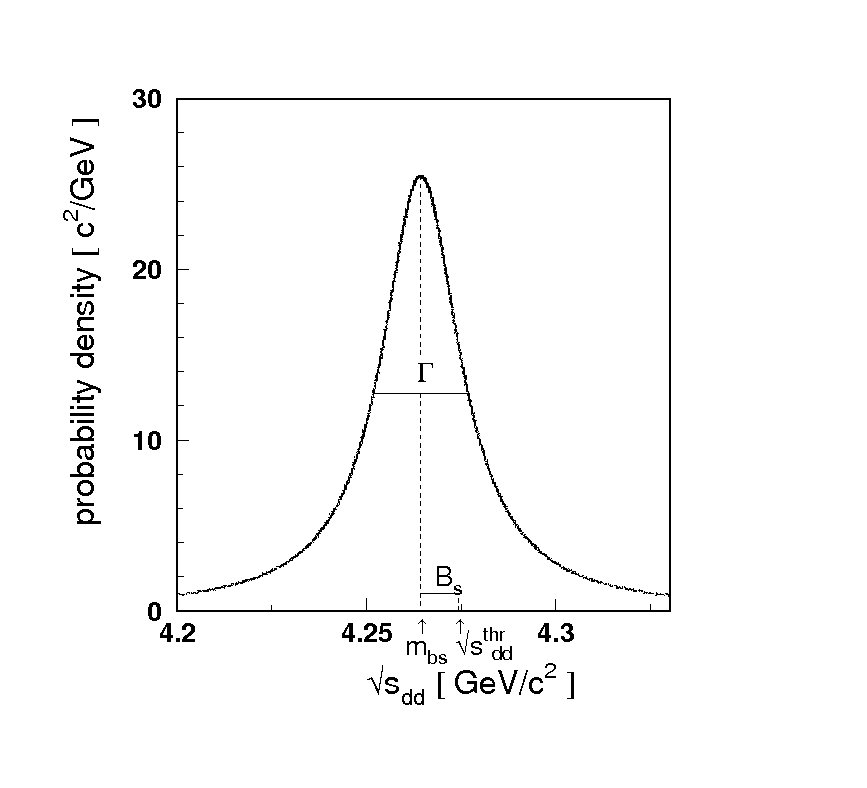}
\caption{Breit-Wigner distribution of square invariant mass $\sqrt{s_{dd}}$.\label{BW}}
\end{figure}

\newpage
\begin{equation}
N\left(\sqrt{s_{dd}}\right)=\frac{1}{2\pi}\frac{\Gamma}{\left(\sqrt{s_{dd}}-m_{bs}\right)^{2}+\Gamma^{2}/4}
\label{eq:BW}
\end{equation}\\

\noindent
where: $m_{bs}=\sqrt{s_{dd}}^{thr}-B_{s}$ - a mass of $\eta$-mesic bound state,\\

$\sqrt{s_{dd}}^{thr}=m_{\eta}+m_{^{4}\hspace{-0.03cm}He}$ - threshold square invariant mass,\\

$B_{s}$ - a binding energy of an $\eta$-mesic bound state,\\

$\Gamma$ - width of an $\eta$-mesic bound state. \\

In the distribution shown in~Fig.~\ref{BW} it is assumed that binding energy equals $B_{s}$=0.01~GeV and a width is equal to 40 MeV what is in agreement with theoretical prediction~\cite{InoueOset}. In simulations a range of $\Gamma$ from 7 to 40 MeV was studied.

\item The $\mbox{N}^{*}$ resonance momentum is distributed isotropically in spherical coordinates of $\eta$-mesic nucleus \mbox{(${p}^{\,\,*}_{F}$, $\theta^{*}$, $\phi^{*}$)} with Fermi momentum distribution of nucleons inside $^{4}\hspace{-0.03cm}\mbox{He}$ which was presented for three different models in Chapter~3. Next, it is transformed into Cartesian coordinates (${\vec{p}}^{\,\,*}_{F}$=(${p}^{\,\,*x}_{F},{p}^{\,\,*y}_{F},{p}^{\,\,*z}_{F}$)) using the following equations:

\begin{equation}
{p}^{\,\,*x}_{F}={p}^{\,\,*}_{F}\cdot sin\theta^* \cdot cos\phi^*
\end{equation}

\begin{equation}
{p}^{\,\,*y}_{F}={p}^{\,\,*}_{F}\cdot sin\theta^* \cdot sin\phi^* 
\end{equation}

\begin{equation}
{p}^{\,\,*z}_{F}={p}^{\,\,*}_{F}\cdot cos\theta^*
\end{equation}  

Here the momentum value ${p}^{\,\,*}_{F}$ is distributed according to the used model, and the direction is simulated isotropically in the space.

\item The $^{3}\hspace{-0.03cm}\mbox{He}$ four momentum vector is calculated (based on spectator model assumption) in the center of mass frame and transformed using Lorentz transformation into laboratory frame. The angle $\theta_{^{3}\hspace{-0.05cm}He}$ is also calculated.

\item Based on $\sqrt{s_{dd}}$ and ${\vec{p}}^{\,\,*}_{F}$ values, resonance mass $m_{{N}^*}$ is calculated according to equation~(\ref{eq:10}).

\item The proton and pion momentum vectors are simulated isotropically in the $\mbox{N}^{*}$ frame in spherical coordinates and transformed into Cartesian coordinates. The absolute value of $\vec{p}^{\,\,**}_{p,\pi^{-}}$ is fixed by equation~(\ref{eq:101}). 

\item The proton and pion four momentum vectors are transformed into the center of mass frame and next into laboratory frame by means of Lorentz transformation. The angles of outgoing proton and pion in LAB system are calculated.

\item The histograms of outgoing particles angle distributions and invariant mass distribution for all generated events are created. 

\item Knowing angles of outgoing particles and the WASA-at-COSY detector geometry it is checked whether generated event can be registered.

\item The histograms of angular distributions and invariant mass distribution of outgoing particles for all accepted events are created.
 
\end{enumerate}

\noindent
The simulation of other three free $\eta$-helium bound states production reactions were carried out in a similar way.\\ 

The simulation of quasi-free reaction, realised with COSY-TOF detection setup is more complex due to the fact that neutron bound in beam deuteron takes part in the reaction of $\eta$-mesic tritium formation. The simulation scheme might be presented as follows:

\begin{enumerate}

\item The beam neutron momentum is distributed isotropically in spherical coordinates with Fermi momentum distribution of nucleons inside beam deuteron (the PARIS and CD-Bonn distributions are presented in Chapter~3) and transformed into Cartesian coordinates ($\vec{p_{n}^{*}}$).

\item The proton spectator four momentum vector as well as its angle with respect to the beam direction are calculated in the beam deuteron center of mass frame and transformed with Lorentz transformation into laboratory frame.

\item The neutron four momentum in the beam deuteron frame and in laboratory frame is calculated using~(\ref{eq:4.2_3}),~(\ref{eq:4.2_4}) and ~(\ref{eq:6.1}),~(\ref{eq:6.2}) formulas, respectively.

\item The square of bound state invariant mass $\sqrt{s_{nd}}$ is calculated based on neutron four momentum and next events are accepted according to the probability described by the Breit-Wigner distribution. 

\item The next points are analogous like in case of free reaction scheme. Moreover, the calculations are carried out for ten values of the deuteron beam momentum $\vec{p}_{beam}$ in the range from 2.6~GeV/c to 3.5~GeV/c.   
  
\end{enumerate}  

\indent
Simulations were conducted for all reactions listed in introduction. Three different values of the bound states width: $\Gamma=\{10,25,40\}$~MeV and three different Fermi momentum distributions of nucleons inside light atomic nuclei were studied. 

\vspace{0.5cm}  
\subsection{Reactions products - angular distributions}

\begin{figure}[h]
\centering
\includegraphics[width=7.1cm,height=6.5cm]{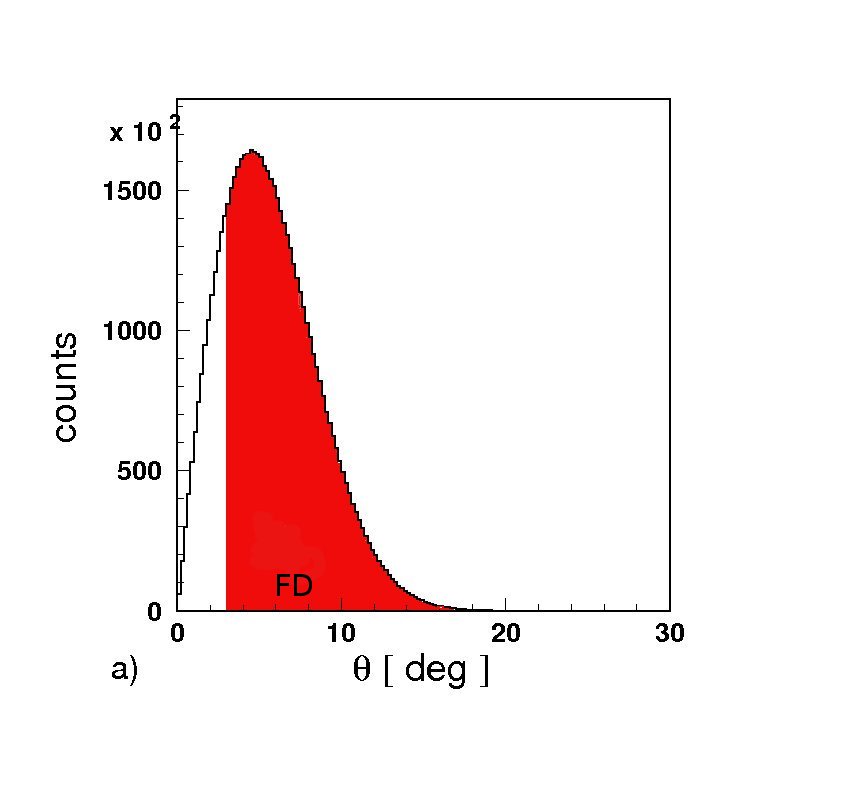} \hspace{-0.5cm}\includegraphics[width=7.1cm,height=6.5cm]{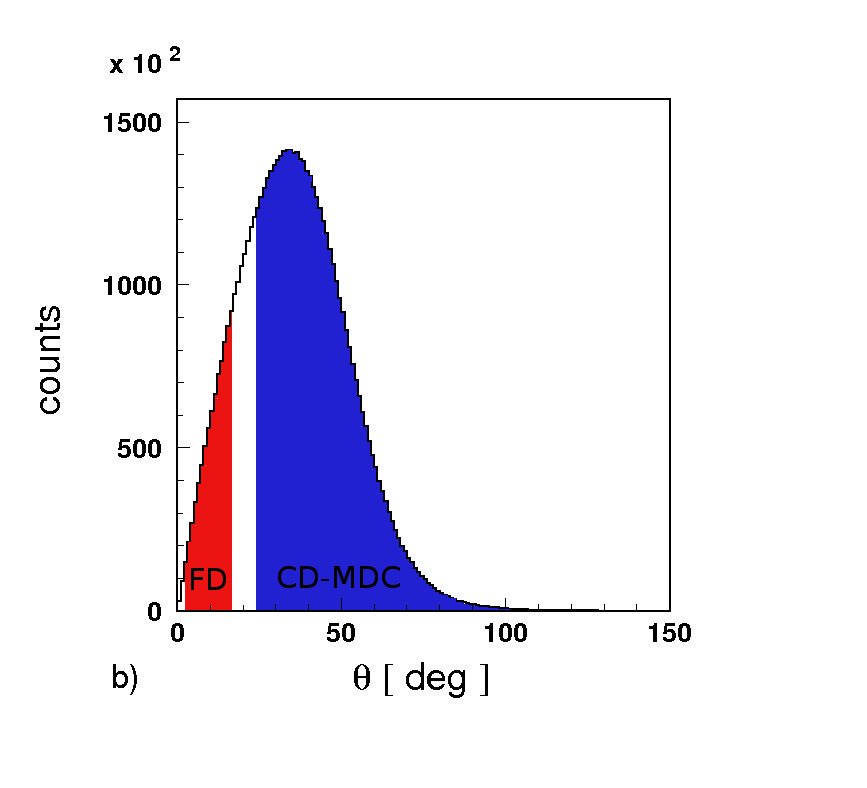} \includegraphics[width=7.1cm,height=6.5cm]{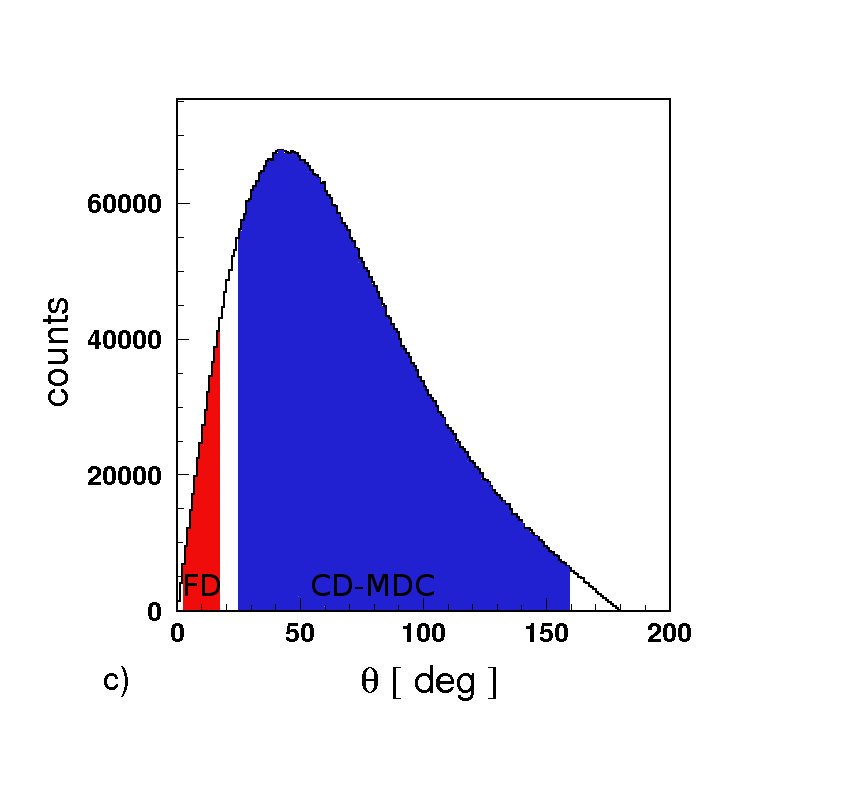}
\caption{Simulated angular distributions of outgoing $^{3}\hspace{-0.05cm}\mbox{He}$ (a), proton (b) and pion (c) formed via \mbox{$dd\rightarrow(^{4}\hspace{-0.03cm}\mbox{He}$-$\eta)_{bs}\rightarrow$ $^{3}\hspace{-0.03cm}\mbox{He} p \pi{}^{-}$} reaction. Figure shows results for $10^{8}$ generated events using the AV18 potential model for the Fermi momentum distribution of nucleons inside $^{4}\hspace{-0.03cm}\mbox{He}$.\label{angular}}
\end{figure}

\noindent
The $\eta$-mesic nuclei, which are formed via deuteron-deuteron, proton-deuteron or neutron-deuteron fussion, decay according to the mechanism which is described in details in Chapter~5.~The distribution of momentum vectors of reaction products depend on the bound state mass distribution and on the distribution of the nucleons momentum inside decayed nuclei. Angular distributions for the outgoing particles being products of reaction (1) are presented in Fig.~\ref{angular}. 

In fact the detectors geometry does not give a possibility to register the whole range of outgoing particles angles. An angular ranges covered by respective components of WASA-at-COSY detection setup are presented in Fig.~\ref{angular} with shaded areas.
 
\vspace{0.5cm}

\subsection{Acceptance}

\noindent
In order to determine acceptance for respective reaction, the studied excess energy range was divided into small intervals.~Next for each interval of Q
\mbox{($\mbox{Q}=\sqrt{s}-\sqrt{s}^{thr}$)} the number of events accepted by detector was divided by number of generated events.
An event is accepted when all outgoing particles being reaction products can be registered in the detector. In simulations all possible combinations of particles registration in different part of detectors were considered. They are presented for particular reactions in Appendix~B. 

\indent
The acceptance for each reaction was determined for three different values of the bound states width $\Gamma$: 10MeV, 25MeV and 40MeV and for three models of nucleons Fermi momentum distributions inside $^{4}\hspace{-0.03cm}\mbox{He}$, $^{3}\hspace{-0.03cm}\mbox{He}$ or T which are in details described in Chapter~3.

\begin{figure}[h!]
\centering
\includegraphics[width=7.5cm,height=7.0cm]{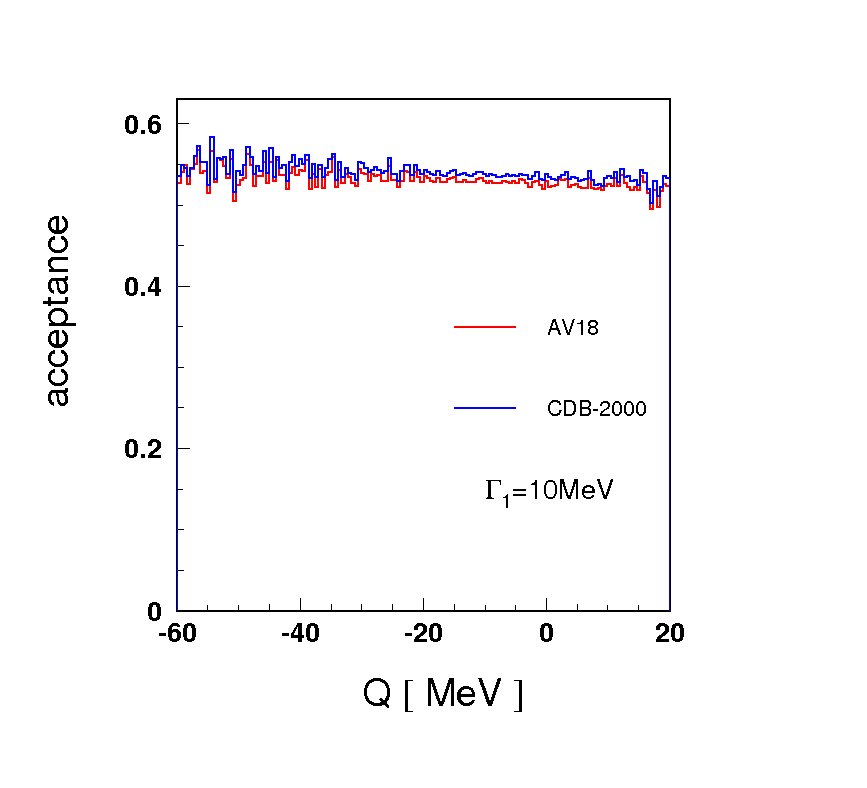} \hspace{-1.7cm} \includegraphics[width=7.5cm,height=7.0cm]{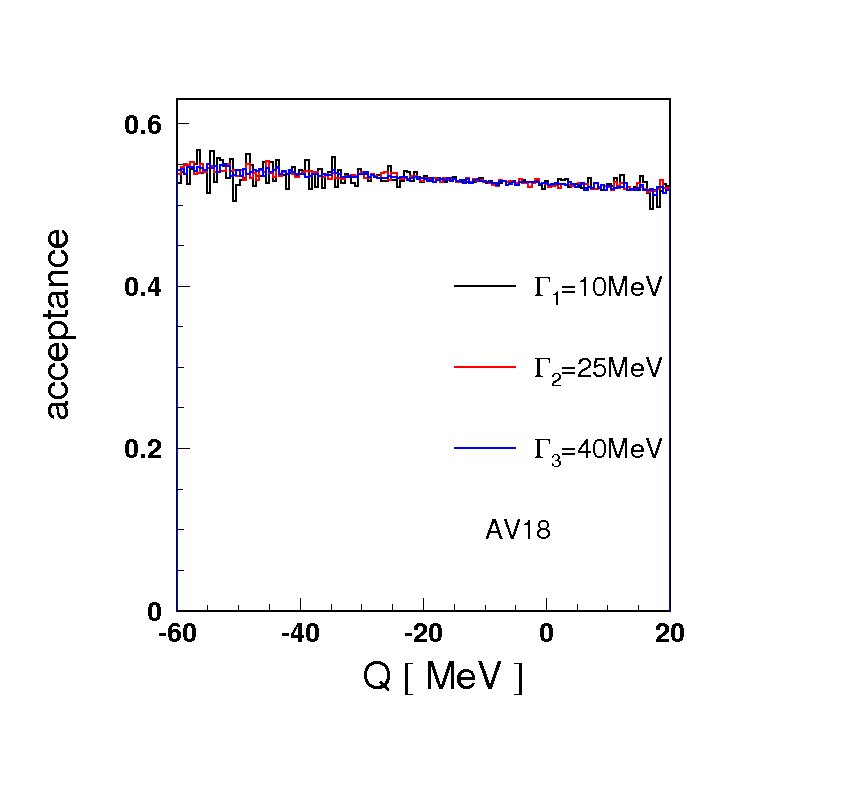}
\vspace{-0.5cm}
\caption{Geometrical acceptances of the WASA-at-COSY detector in case of $dd\rightarrow(^{4}\hspace{-0.03cm}\mbox{He}$-$\eta)_{bs}\rightarrow$ $^{3}\hspace{-0.03cm}\mbox{He} p \pi{}^{-}$ reaction for the $\Gamma_{1}$=10 MeV width and AV18 and CDB-2000 models of nucleon Fermi momentum distribution inside $^{4}\hspace{-0.03cm}\mbox{He}$ (left) as well as for AV18 Fermi momentum distribution model and three different values of $\Gamma$ (right) \label{reakcja_1}.}
\end{figure}

\begin{figure}[h!]
\centering
\includegraphics[width=7.5cm,height=7.0cm]{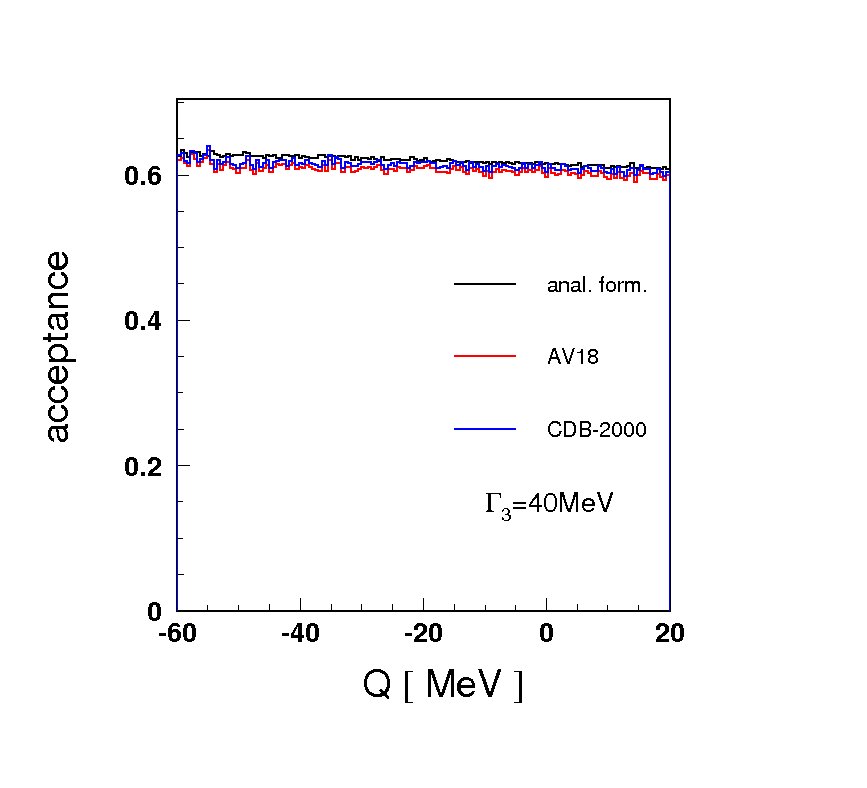} \hspace{-1.7cm} \includegraphics[width=7.5cm,height=7.0cm]{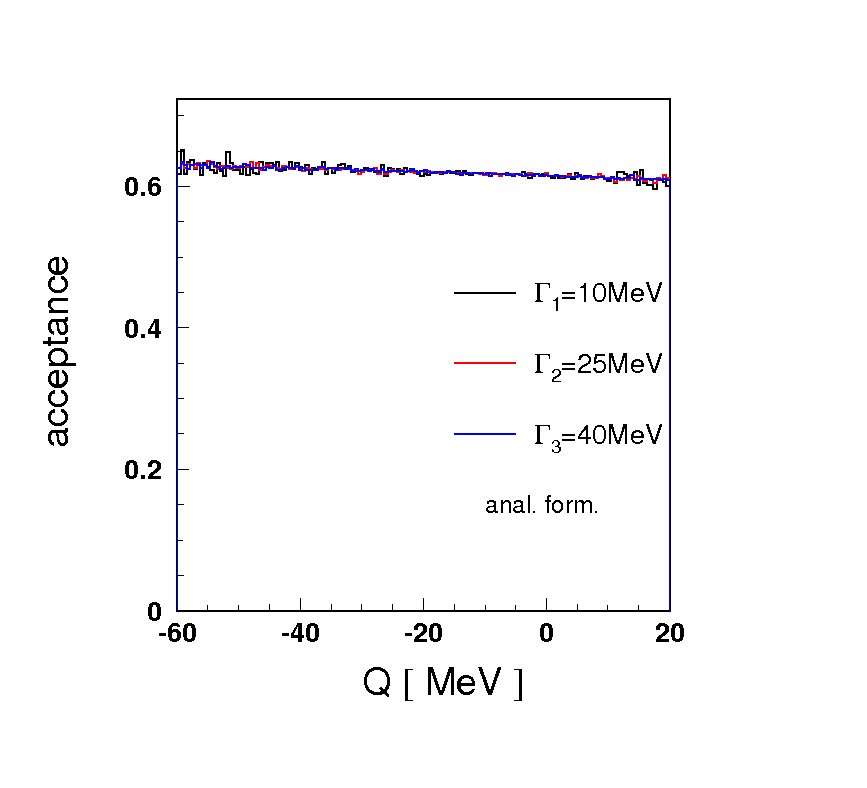}
\vspace{-0.5cm}
\caption{Geometrical acceptances in case of \mbox{$pd\rightarrow(^{3}\hspace{-0.03cm}\mbox{He}$-$\eta)_{bs}\rightarrow$ $d p \pi{}^{0}\rightarrow d p \gamma \gamma$} reaction for the $\Gamma_{3}$=40 MeV width and three different models of nucleon Fermi momentum distribution inside $^{3}\hspace{-0.03cm}\mbox{He}$ (left) as well as for analitic formula describing nucleon Fermi momentum distribution and three different values of $\Gamma$ (right).}\label{reakcja_3}
\end{figure}

\indent
The WASA acceptance as a function of the excess energy Q is presented for two reactions of free $^{4}\hspace{-0.03cm}\mbox{He}$-$\eta$ and $^{3}\hspace{-0.03cm}\mbox{He}$-$\eta$ production: $dd\rightarrow(^{4}\hspace{-0.03cm}\mbox{He}$-$\eta)_{bs}\rightarrow$ $^{3}\hspace{-0.03cm}\mbox{He} p \pi{}^{-}$ and $pd\rightarrow(^{3}\hspace{-0.03cm}\mbox{He}$-$\eta)_{bs}\rightarrow$ $d p \pi{}^{0}\rightarrow$ $d p \gamma \gamma$ in Fig.~\ref{reakcja_1} and  Fig.~\ref{reakcja_3}, respectively.
The acceptance is almost constant as a function of excess energy and it is independent of the value of $\Gamma$ and model of Fermi momentum distribution within a few per cent.
For two other reactions situation is analogous.

\newpage

\indent
The average values of WASA geometrical acceptances for registering the considered free reactions were calculated using following formula:
\begin{equation}
A=\frac{\sum_{Q}\frac{N_{acc}(Q)}{N_{gen}(Q)}}{N}
\label{eq:A}
\end{equation}

\noindent
where:  \hspace{0.2cm} $N_{acc}(Q)$- the number of accepted events in a given interval of Q,\\

 \hspace{1.0cm}  $N_{gen}(Q)$- the number of generated events in a given interval of Q,\\
 
 \hspace{1.0cm}  $N$- the number of all ranges of summation. \\
  
\noindent
The obtained results for different gamma values and different models of nucleon momentum distributions are presented in following tables:

\vspace{-0.5cm}

\noindent
\begin{table}[h!]
\begin{footnotesize}
\begin{center}
\begin{tabular}{|c|c|c|c|}\hline
[MeV/c] &AV18 &CDB2000\\
\hline 
$\Gamma1$  &0.5296 &0.5380 \\
$\Gamma2$  &0.5301 &0.5377 \\
$\Gamma3$  &0.5323 &0.5402 \\
\hline
\end{tabular}
\end{center}
\begin{center}
\caption{Average acceptances of WASA-at-COSY detector for the $dd\rightarrow(^{4}\hspace{-0.03cm}\mbox{He}$-$\eta)_{bs}\rightarrow$ $^{3}\hspace{-0.03cm}\mbox{He} p \pi{}^{-}$ reaction for three different $\Gamma$ values and AV18 and CDB-2000 models of nucleon momentum distribution inside $^{4}$\hspace{-0.03cm}$\mbox{He}$.}
\end{center}
\end{footnotesize}
\end{table}

\vspace{-1.4cm}

\noindent
\begin{table}[h!]
\begin{footnotesize}
\begin{center}
\begin{tabular}{|c|c|c|c|}\hline
[MeV/c]  &AV18 &CDB2000\\
\hline 
$\Gamma1$  &0.3785 &0.3889 \\
$\Gamma2$  &0.3758 &0.3877 \\
$\Gamma3$  &0.3749 &0.3866 \\
\hline
\end{tabular}
\end{center}
\begin{center}
\caption{Average acceptances of WASA-at-COSY detector for the $dd\rightarrow(^{4}\hspace{-0.03cm}\mbox{He}$-$\eta)_{bs}\rightarrow$ $d p p \pi{}^{-}$ reaction for three different $\Gamma$ values and AV18 and CDB-2000 models of nucleon momentum distribution inside $^{4}\hspace{-0.03cm}\mbox{He}$.}
\end{center}
\end{footnotesize}
\end{table}

\vspace{-1.4cm}

\noindent
\begin{table}[h!]
\begin{footnotesize}
\begin{center}
\begin{tabular}{|c|c|c|c|}\hline
[MeV/c]  &anal. form. &AV18 &CDB2000\\
\hline 
$\Gamma1$  &0.6175 &0.6071 &0.6139 \\
$\Gamma2$  &0.6187 &0.6091 &0.6130 \\
$\Gamma3$  &0.6188 &0.6067 &0.6135 \\
\hline
\end{tabular}
\end{center}
\begin{center}
\caption{Average acceptances of WASA-at-COSY detector for the $pd\rightarrow(^{3}\hspace{-0.03cm}\mbox{He}$-$\eta)_{bs}\rightarrow$ $d p \pi{}^{0}\rightarrow d p \gamma \gamma$ reaction for three different $\Gamma$ values and three different models of nucleon momentum distribution inside $^{3}\hspace{-0.03cm}\mbox{He}$.}
\end{center}
\end{footnotesize}
\end{table}

\vspace{-1.4cm}

\newpage
\noindent
\begin{table}[h!]
\begin{footnotesize}
\begin{center}
\begin{tabular}{|c|c|c|c|}\hline
[MeV/c]  &anal. form. &AV18 &CDB2000\\
\hline 
$\Gamma1$  &0.5054 &0.5059 &0.5052 \\
$\Gamma2$  &0.5083 &0.5029 &0.5042 \\
$\Gamma3$  &0.5054 &0.5070 &0.5045 \\
\hline
\end{tabular}
\end{center}
\begin{center}
\caption{Average acceptances of WASA-at-COSY detector for the $pd\rightarrow(^{3}\hspace{-0.03cm}\mbox{He}$-$\eta)_{bs}\rightarrow$ $p p p \pi{}^{-}$ reaction for three different $\Gamma$ values and three different models of nucleon momentum distribution inside $^{3}\hspace{-0.03cm}\mbox{He}$.}
\end{center}
\end{footnotesize}
\end{table}

\vspace{-1.0cm}

\begin{figure}[h!]
\centering
\includegraphics[width=8.0cm,height=7.5cm]{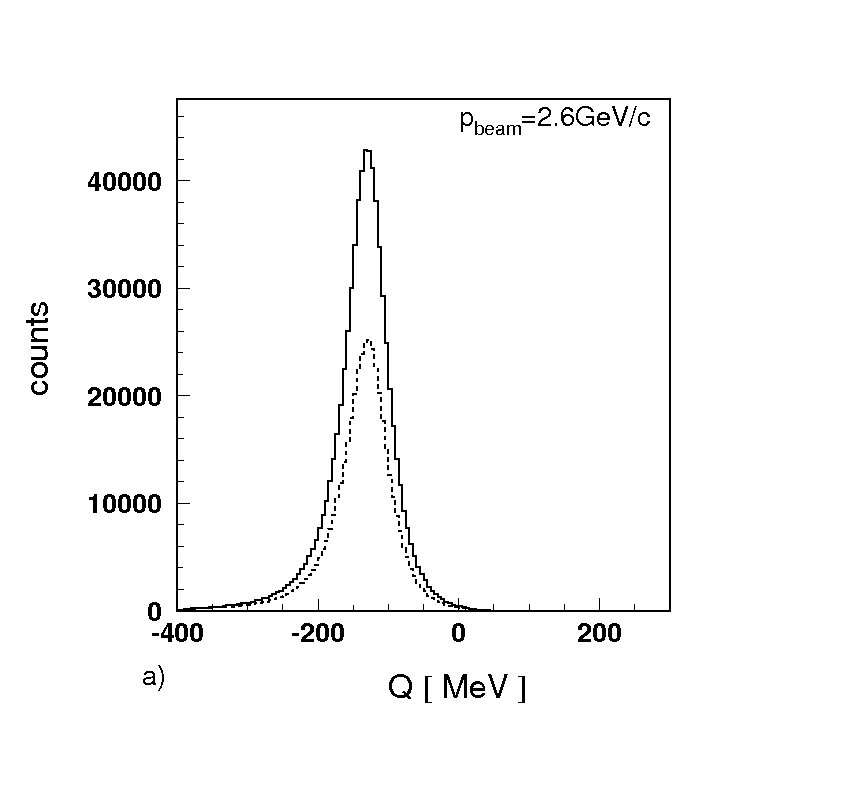} \hspace{-1.9cm} \includegraphics[width=8.0cm,height=7.5cm]{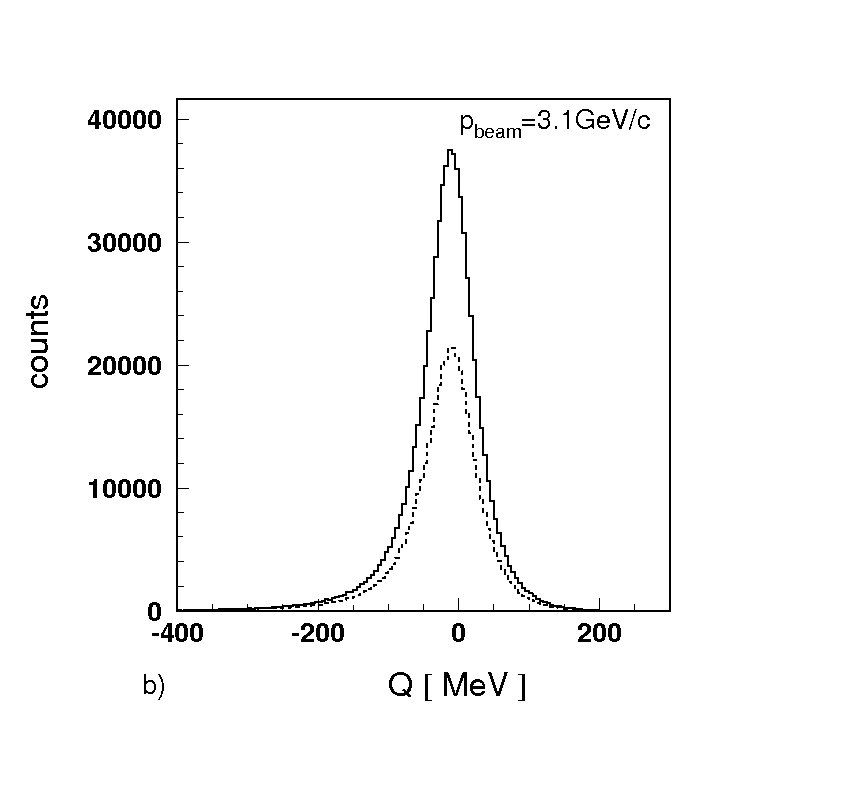} \includegraphics[width=8.0cm,height=7.5cm]{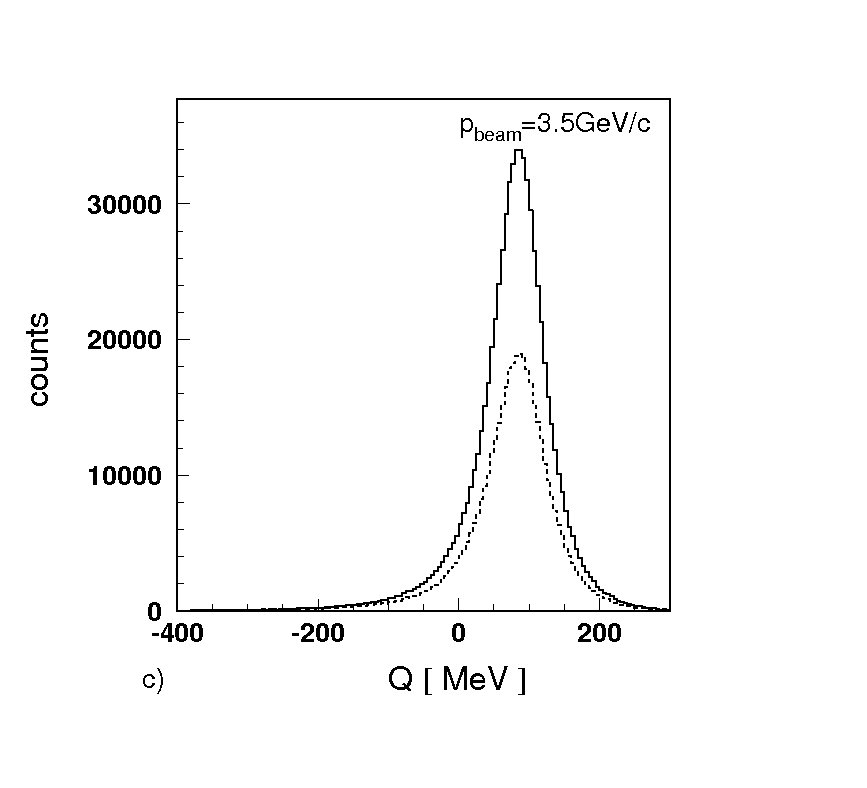}
\caption{The number of generated (solid) and accepted \mbox{$nd\rightarrow(\mbox{T}$-$\eta)_{bs}\rightarrow$ $d p \pi{}^{-}$} events (dashed) as a function of excess energy Q with respect to quasi-free $nd\rightarrow(\mbox{T}$-$\eta)_{bs}$ reaction at deuteron beam momentum ${p}_{beam}$=2.6 GeV/c (a), \mbox{${p}_{beam}$=3.1 GeV/c (b)} and \mbox{${p}_{beam}$=3.5 GeV/c} (c).\label{Q}}
\end{figure}

\indent
For the quasi-free reaction $dd\rightarrow p_{sp}(\mbox{T}$-$\eta)_{bs}\rightarrow$ $ p_{sp}d p \pi{}^{-}$ the COSY-TOF acceptance was calculated for ten of beam momentum values for the range of $\vec{p}_{beam}$=2.6~GeV/c to 3.5~GeV/c. The excess energy distribution for \mbox{$nd\rightarrow(\mbox{T}$-$\eta)_{bs}$} reactions determined by the Fermi momentum distribution of neutron inside deuteron is presented in Fig.~\ref{Q} for ${p}_{beam}$ equal to 2.6 GeV/c, 3.1 GeV/c and 3.5 GeV/c. 

For each case, the number of $nd\rightarrow(\mbox{T}$-$\eta)_{bs}$ generated events as a function of the excess energy Q is represented by solid line while the excess energy distribution for accepted events is shown by dashed line. The threshold deuteron beam momentum for the $dd\rightarrow p_{sp}(\mbox{T}$-$\eta)_{bs}$ is equal 3.1 GeV/c for the case if the Fermi momentum in the beam deuteron is equal to 0. 

\begin{figure}[h!]
\centering
\includegraphics[width=8.0cm,height=7.5cm]{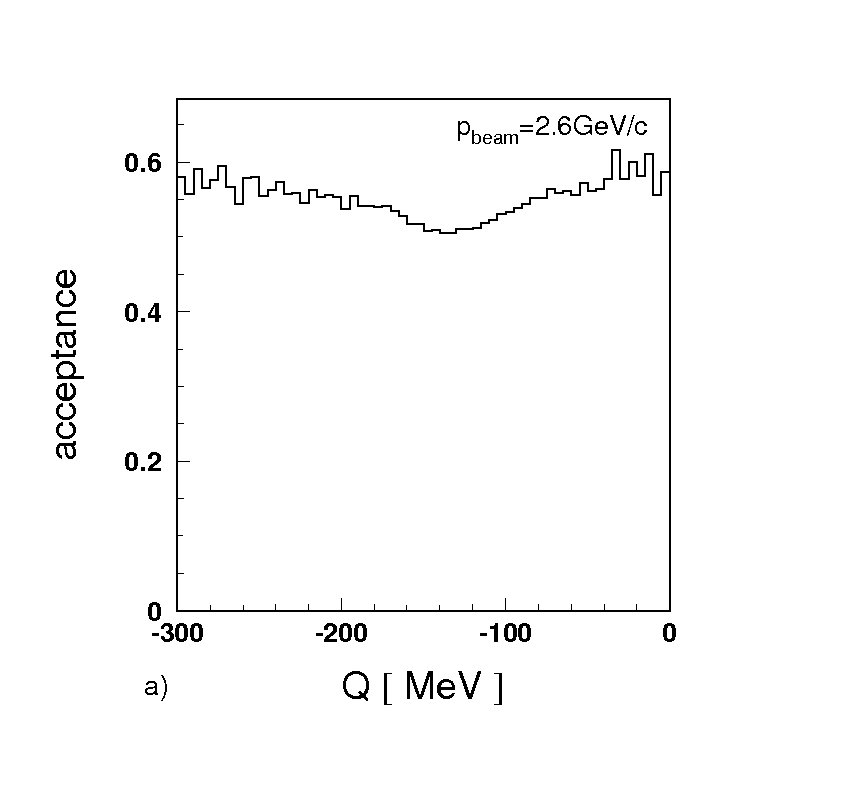} \hspace{-1.7cm}\includegraphics[width=8.0cm,height=7.5cm]{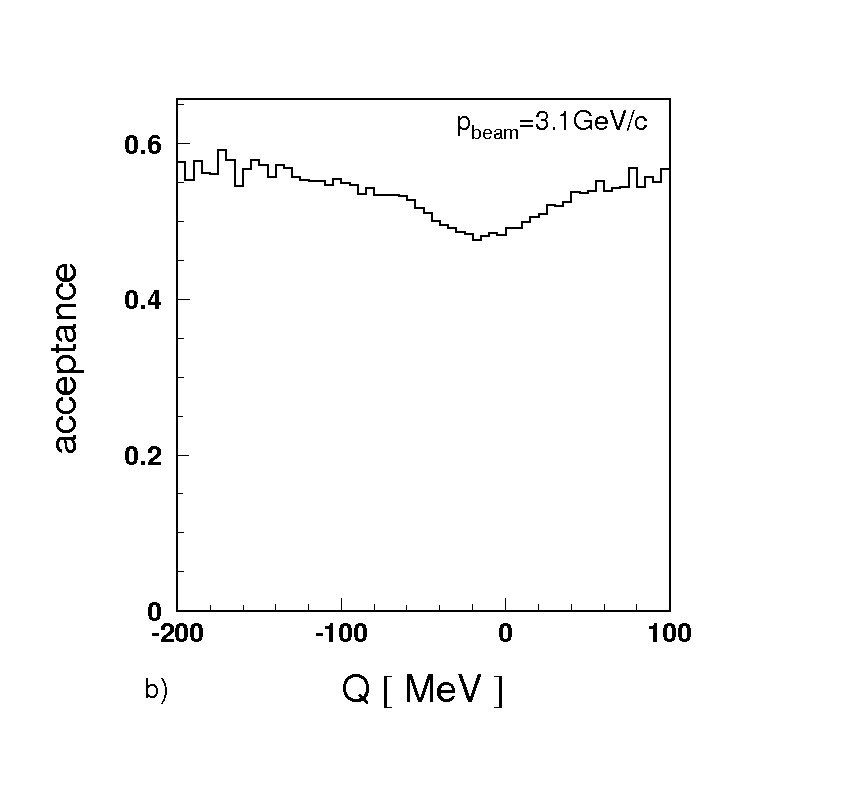} \includegraphics[width=8.0cm,height=7.5cm]{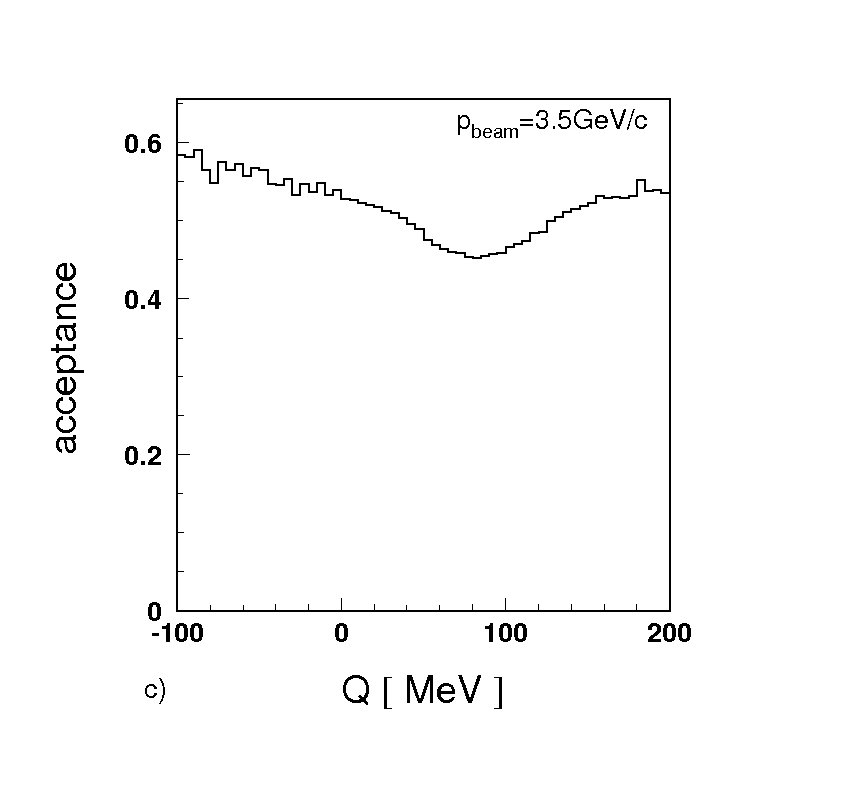} \hspace{-1.7cm}\includegraphics[width=8.1cm,height=7.8cm]{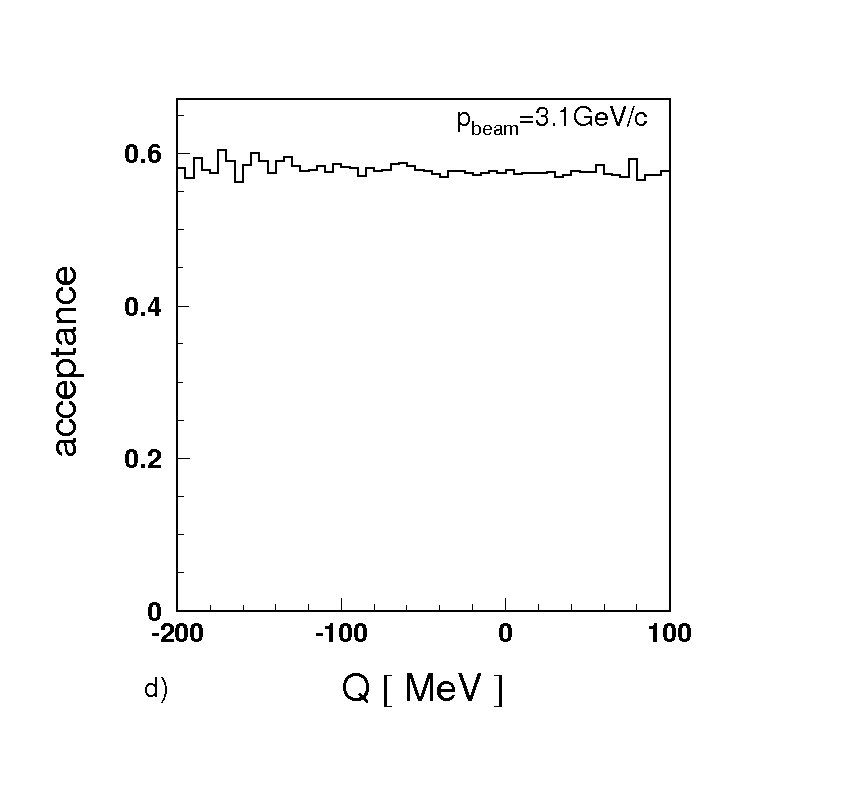}
\caption{Geometrical COSY-TOF acceptances in case of the $dd\rightarrow p_{sp}(\mbox{T}$-$\eta)_{bs}\rightarrow$ $ p_{sp}d p \pi{}^{-}$ reaction at ${p}_{beam}$=2.6 GeV/c (a), ${p}_{beam}$=3.1 GeV/c (b) and ${p}_{beam}$=3.5 GeV/c (c). For the simulations Paris model of nucleon Fermi momentum distribution inside deuteron and $\Gamma_{1}$=10 MeV bound state width and analitic formula describing nucleon Fermi momentum distribution inside T were used. Panel (d) represents the reaction acceptance for \mbox{${p}_{beam}$=3.1 GeV/c} with assumption that the detector acceptance for proton spectator registration equals 1.\label{pbeam}}
\end{figure}

\indent
Similarly like in case of free reactions, the COSY-TOF acceptance function is obtained by dividing the number of accepted events by the number of generated events for each of excess energy intervals. Respective results for three values of deuteron beam momentum are presented in Fig.~\ref{pbeam} (a), (b) and (c). 

The acceptance of ejectiles registration in the quasi-free reaction is not constant function of excess energy. It decreases for Q values corresponding to the Fermi momentum of neutron in the deuteron beam equal to zero and results from the fact that proton spectator is not accepted by detector geometry if its Fermi momentum value in deuteron frame equals $p^{*}_{sp}$=0. For a comparison, Fig.~\ref{pbeam} (d) presents that the acceptance dependence of Q would be constant if the proton spectator was accepted. 

\newpage
In order to determine the effective acceptance for registering the quasi-free \mbox{$dd\rightarrow p_{sp}(\mbox{T}$-$\eta)_{bs}\rightarrow$ $ p_{sp}d p \pi{}^{-}$} reaction in the near to threshold Q range from -60~MeV to 20 MeV when using the TOF detector the following formula is used:

\begin{equation}
A_{eff}=A\frac {N_{(-60,20)}}{N_{gen}}
\end{equation}

\noindent
where:  \hspace{0.2cm} $A$- the average acceptance given by formula~(\ref{eq:A}),\\

 \hspace{1.0cm}  $N_{gen}$- the number of generated events,\\
 
 \hspace{1.0cm}  $N_{(-60,20)}$- the number of generated events for Q$\in$(-60,20)~MeV.\\

\begin{figure}[h!]
\centering
\includegraphics[width=10.0cm,height=9.0cm]{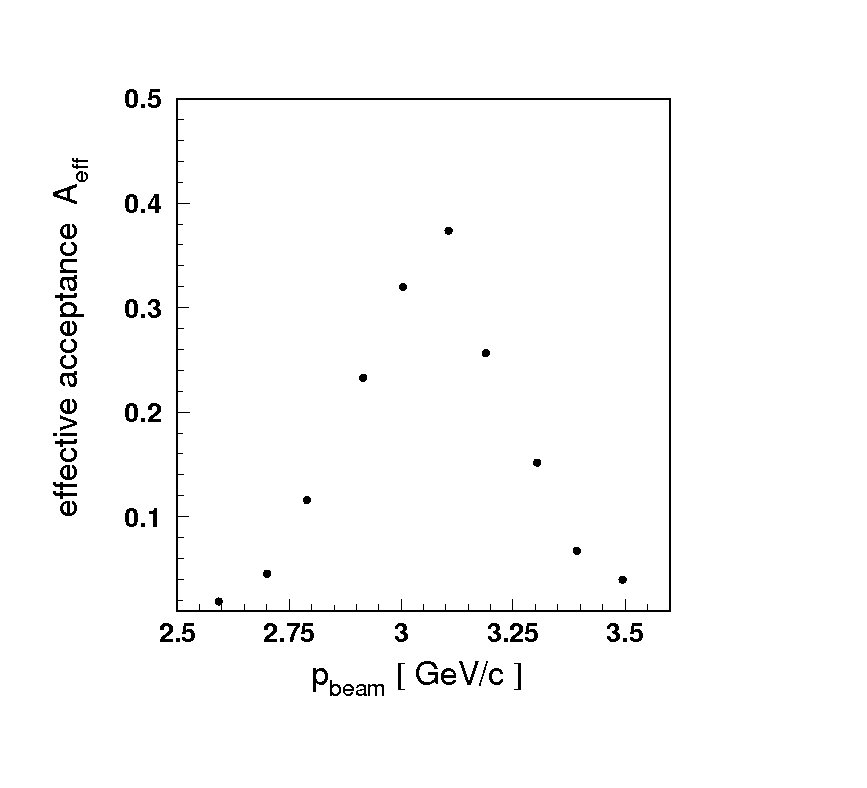} 
\caption{Effective acceptance for the registration of the quasi-free \mbox{$dd\rightarrow p_{sp}(\mbox{T}$-$\eta)_{bs}\rightarrow$ $ p_{sp}d p \pi{}^{-}$} reaction near the $\eta$ production threshold (Q$\in$(-60,20)~MeV) as a function of beam momentum.\label{Pbeam}}
\end{figure}

The effective acceptance $A_{eff}$ was calculated for ten values of the beam momentum. The dependence $A_{eff}(p_{beam})$ is presented in Fig.~\ref{Pbeam}. 

In the simulations of the \mbox{$dd\rightarrow p_{sp}(\mbox{T}$-$\eta)_{bs}\rightarrow$ $ p_{sp}d p \pi{}^{-}$} reaction it is assumed that the square of invariant mass of T-$\eta$ nuclei is given by Breit-Wigner distribution. In this case an effective acceptance for the registration of the bound state decay products is calculated by formula:

\begin{equation}
A^{BW}_{eff}=A\frac {N^{BW}_{(-60,20)}}{N^{BW}_{gen}}
\end{equation}

\noindent
where:  \hspace{0.2cm} $A$- the average acceptance given by formula~(\ref{eq:A}),\\

 \hspace{1.0cm}  $N^{BW}_{gen}$- the number of generated events accepted with probability
  
 \hspace{1.0cm} calculated according to the Breit-Wigner distribution of $\sqrt{s_{nd}}$,\\
 
 \hspace{1.0cm}  $N^{BW}_{(-60,20)}$- the number of generated events for Q$\in$(-60,20)~MeV 
 
 \hspace{1.0cm} accepted with probability calculated according to the Breit-Wigner 
    
\begin{figure}[h!]
\centering
\includegraphics[width=10.0cm,height=9.0cm]{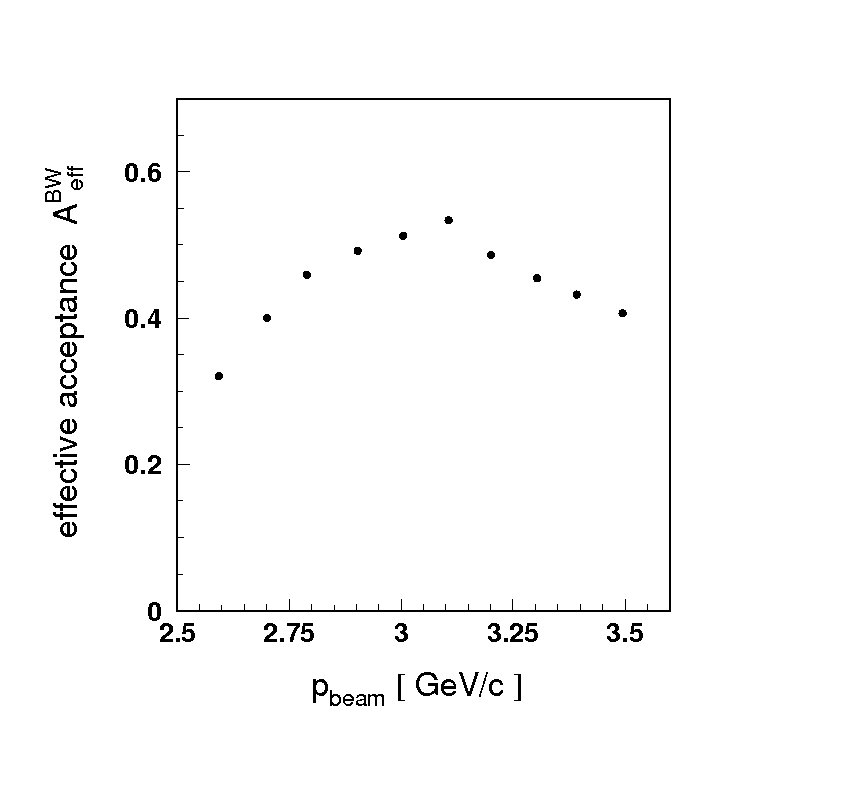}
\caption{Effective acceptance for the registration of the quasi free \mbox{$dd\rightarrow p_{sp}(\mbox{T}$-$\eta)_{bs}\rightarrow$ $ p_{sp}d p \pi{}^{-}$}  reaction near the $\eta$ production threshold (Q$\in$(-60,20)~MeV) as a function of beam momentum assuming a Breit-Wigner distribution of $\sqrt{s_{nd}}$ for the bound state width $\Gamma$=10MeV. \label{Pbeam1}}
\end{figure}

 \hspace{1.0cm} distribution of $\sqrt{s_{nd}}$\\
    
\noindent 
and is presented in Fig.~\ref{Pbeam1} as a function of the beam deuteron momentum.

From the dependence shown in above figure it results that the highest probability for the registration of considered quasi-free \mbox{$dd\rightarrow p_{sp}(p_{F}=0)(\mbox{T}$-$\eta)_{bs}\rightarrow$ $ p_{sp}d p \pi{}^{-}$} reaction is for the beam momentum $p_{beam}$=3.1~GeV/c corresponding to the \mbox{$dd\rightarrow p_{sp}(\mbox{T}$-$\eta)_{bs}\rightarrow$ $ p_{sp}d p \pi{}^{-}$}  reaction threshold whereas for the beam momentum above and below the threshold the effective acceptance decreases.  
Above considerations were carried out for the Paris model of nucleon Fermi momentum distribution inside deuteron, $\Gamma_{1}$=10~MeV bound state width and analitic formula describing nucleon Fermi momentum distribution inside T. The calculations are consistent with the one for CD-Bonn model of nucleon momentum distribution inside deuteron with an accurancy better than 4\% and are independent of models of nucleon Fermi momentum distribution inside tritium.


\newpage
\pagestyle{fancy}
\fancyhf{} 
\fancyhead[LE,RO]{\textbf{\thepage}}
\fancyhead[RE]{\small\textbf{{Summary and conclusions}}} 

\newpage
\thispagestyle{plain}
\section {Summary and conclusions}

\vspace{0.5cm}

\noindent
The main aim of this thesis was to test the feasibility of the measurement of $\eta$-helium mesic nucleus production via reactions:\\

\noindent
$dd\rightarrow(^{4}\hspace{-0.03cm}\mbox{He}$-$\eta)_{bs}\rightarrow$ $^{3}\hspace{-0.03cm}\mbox{He} p \pi{}^{-}$\\
$dd\rightarrow(^{4}\hspace{-0.03cm}\mbox{He}$-$\eta)_{bs}\rightarrow d p p \pi{}^{-}$\\
$pd\rightarrow(^{3}\hspace{-0.03cm}\mbox{He}$-$\eta)_{bs}\rightarrow d p \pi{}^{0} \rightarrow d p \gamma \gamma$\\ 
$pd\rightarrow(^{3}\hspace{-0.03cm}\mbox{He}$-$\eta)_{bs}\rightarrow p p p \pi{}^{-}$,\\

\noindent
with WASA-at-COSY detector and test the feasibility of the $\eta$-T bound state production via reaction:\\ 

\noindent
$nd\rightarrow(\mbox{T}$-$\eta)_{bs}\rightarrow d p \pi{}^{-}$,\\

\noindent
realised through \mbox{$dd\rightarrow p_{sp}(\mbox{T}$-$\eta)_{bs}\rightarrow$ $ p_{sp}d p \pi{}^{-}$} quasi-free reaction with \mbox{COSY-TOF} detection setup.

\indent
The reaction kinematics for free and quasi-free $\eta$-mesic bound states production as well as the spectator model assumptions were analysed and discussed. Moreover, the nucleon momentum distribution inside $^{4}\hspace{-0.03cm}\mbox{He}$, $^{3}\hspace{-0.03cm}\mbox{He}$, T and d, was presented for different models based on the analytic formulas and theoretical analysis for each of nuclei. Numerical data for those distributions are presented in Appendix~A.
The simulation were carried out and acceptances of \mbox{WASA-at-COSY} detection systems for free processes were calculated and compared for three assumed values of bound states width as well as for three different models of nucleon Fermi momentum distribution. In case of quasi-free reaction the effective \mbox{COSY-TOF} acceptance as a function of beam momentum was determined.  
 
\indent
The simulation results show that the acceptance as function of excess energy for the free reactions of $\eta$-helium production is in good approximation constant near the $\eta$ production threshold and is independent of the bound state width value and model of Fermi momentum distribution. The calculations present that the absolute value of acceptance depends on the reaction channel. The most probable is registering of particles outgoing from the \mbox{$pd\rightarrow(^{3}\hspace{-0.03cm}\mbox{He}$-$\eta)_{bs}\rightarrow d p \pi{}^{0} \rightarrow d p \gamma \gamma$} reaction. The average value of WASA geometrical acceptance is then equal to A=0.61. 

\indent
For the quasi-free reaction $dd\rightarrow p_{sp}(\mbox{T}$-$\eta)_{bs}\rightarrow$ $ p_{sp}d p \pi{}^{-}$ of $\eta$-tritium formation the \mbox{COSY-TOF} effective acceptance is dependent on the beam momentum value. It reaches the highest value for the beam momentum of $p_{beam}$=3.1~GeV/c corresponding to the \mbox{$dd\rightarrow p_{sp}(p_{F}=0)n d\rightarrow p_{sp}(p_{F}=0)\mbox{T}$-$\eta$} reaction threshold. Thus the measurement of quasi-free reaction products is the most efficient at the $\eta$ production threshold. 

\indent
In this thesis the geometrical acceptance for particular reactions was calculated. In the future, the efficiency of registration for the outgoing particles in each of reaction will be determined. \\
  
\indent 
Finally the experiments in which $\eta$-nuclei will be searched are interesting because of following issues:

\begin{enumerate} 
\item Potential of discovering of $\eta$-mesic bound states. Observation of that state would allow to investigate interactions between the $\eta$ meson and the nucleons inside a nuclear matter.
\item $\eta$-mesic bound systems would provide the information about $N^{*}$(1535)~\cite{Jido} resonance and $\eta$ meson properties in nuclear matter~\cite{InoueOset}.
\item The existence of the bound states could give a possibility to study a flavour singlet component of the quark-gluon wave function of the $\eta$ meson~\cite{BassTom, BassThomas}.
\end{enumerate}

\indent
The simulations presented in this thesis show that the $\eta$-helium nuclei measurement can be carried out at WASA-at-COSY detection setup, while the $\eta$-tritium bound states might be searched at COSY-TOF detector. As a result of simulation it is established that the most efficient measurement of \mbox{$dd\rightarrow p_{sp}(\mbox{T}$-$\eta)_{bs}\rightarrow$ $ p_{sp}d p \pi{}^{-}$} reaction at COSY-TOF detector can be done at the beam momentum of 3.1~GeV/c.  


\newpage
\thispagestyle{plain}

\normalsize

\appendix 
  \section{Fermi momentum distributions-numerical data}
  
\vspace{0.5cm}  
  
\pagestyle{fancy}
\fancyhf{}
\fancyhead[LE,RO]{\textbf{\thepage}}
\fancyhead[RE]{\small\textbf{{Appendix}}} 
 
\noindent
In Chapter 3 Fermi momentum distributions of nucleons inside d, T, $^{3}\mbox{He}$ and $^{4}\mbox{He}$ nuclei were described and presented in Fig.~\ref{deuteron}, \ref{hel_3}, \ref{hel_4}. Numerical data for these distributions are given in the following tables:

\begin{table}[h!]
\scriptsize
\begin{center}
\begin{tabular}{|c|c|c|}\hline
$p_{F}$ &\multicolumn{2}{|c|}{$f(p_F)$ [c/MeV]}\\
\cline{2-3} 
[MeV/c] &Paris~\cite{Lacombe} &CDBonn~\cite{Machleidt}\\
\hline
10  &0.00218 &0.00195 \\
20  &0.00662 &0.00581 \\
30  &0.00997 &0.00894 \\
40  &0.01138 &0.01028 \\
50  &0.01126 &0.00999 \\
60  &0.01017 &0.00975 \\
70  &0.00881 &0.00803 \\
80  &0.00742 &0.00685 \\
90  &0.00624 &0.00582 \\
100 &0.00523 &0.00483 \\
110 &0.00427 &0.00407 \\
120 &0.00357 &0.00336 \\
130 &0.00301 &0.00284 \\
140 &0.00241 &0.00236 \\
150 &0.00194 &0.00199 \\
160 &0.00162 &0.00167 \\
170 &0.00138 &0.00142 \\
180 &0.00116 &0.00119 \\
190 &0.00096 &0.00102 \\
200 &0.00080 &0.00087 \\
210 &0.00071 &0.00077 \\
220 &0.00064 &0.00067 \\
230 &0.00054 &0.00057 \\
240 &0.00046 &0.00048 \\
250 &0.00044 &0.00043 \\
260 &0.00037 &0.00038 \\
270 &0.00032 &0.00032 \\
280 &0.00029 &0.00028 \\
290 &0.00027 &0.00025 \\
300 &0.00024 &0.00023 \\
310 &0.00021 &0.00020 \\
320 &0.00019 &0.00018 \\
330 &0.00016 &0.00016 \\
340 &0.00014 &0.00015 \\
350 &0.00013 &0.00013 \\
360 &0.00012 &0.00012 \\
370 &0.00010 &0.00011 \\
380 &0.00010 &0.00010 \\
390 &0.00008 &0.00009 \\
400 &0.00008 &0.00008 \\ \hline
\end{tabular}
\end{center}
\begin{center}
\caption{Fermi momentum distribution of nucleon inside deuteron according to Paris model and CDBonn model (Chapter 3). The distributions were normalized to unity in the momentum range from 0 to 400 MeV/c.}
\end{center}
\end{table}

\newpage
\begin{table}[h!]
\footnotesize
\begin{center}
\begin{tabular}{|c|c|c|c|}\hline
$p_{F}$ &\multicolumn{3}{|c|}{$f(p_F)$ [c/GeV]}\\
\cline{2-4} 
[Gev/c] &anal. formula~\cite{Abdullin} &AV18+UrbIX~\cite{Nogga1,Nogga2} &CDB2000+TM~\cite{Nogga1,Nogga2} \\
\hline
0.01 &0.4788 &0.4896 &0.4877 \\
0.02 &1.7861 &1.7752 &1.7681 \\
0.03 &3.5812 &3.4555 &3.4418 \\
0.04 &5.4299 &5.1298 &5.1083 \\
0.05 &6.9442 &6.4313 &6.4083 \\
0.06 &7.8847 &7.2795 &7.2566 \\
0.07 &8.1972 &7.6620 &7.4559 \\
0.08 &7.9806 &7.6485 &7.6352 \\
0.09 &7.4157 &7.3449 &7.3401 \\
0.10 &6.6911 &6.8532 &6.8579 \\
0.11 &5.9512 &6.2508 &6.2654 \\
0.12 &5.2771 &5.5925 &5.6167 \\
0.13 &4.6944 &4.9449 &4.9778 \\
0.14 &4.1926 &4.3187 &4.3593 \\
0.15 &3.7476 &3.7398 &3.7862 \\
0.16 &3.3361 &3.2234 &3.2754 \\
0.17 &2.9434 &2.7373 &2.7934 \\
0.18 &2.5639 &2.3459 &2.4038 \\
0.19 &2.1996 &1.9769 &2.0363 \\
0.20 &1.8557 &1.6787 &1.7378 \\
0.21 &1.5384 &1.4070 &1.4653 \\
0.22 &1.2531 &1.1897 &1.2459 \\
0.23 &1.0029 &0.9913 &1.0451 \\
0.24 &0.7887 &0.8398 &0.8901 \\
0.25 &0.6096 &0.6943 &0.7410 \\
0.26 &0.4633 &0.5935 &0.6358 \\
0.27 &0.3462 &0.4927 &0.5306 \\
0.28 &0.2545 &0.4219 &0.4550 \\
0.29 &0.1840 &0.3579 &0.3860 \\
0.30 &0.1309 &0.3036 &0.3269 \\
0.31 &0.0916 &0.2650 &0.2833 \\
0.32 &0.0631 &0.2264 &0.2398 \\
0.33 &0.0428 &0.2014 &0.2101 \\
0.34 &0.0286 &0.1794 &0.1834 \\
0.35 &0.0188 &0.1587 &0.1581 \\
0.36 &0.0121 &0.1469 &0.1422 \\
0.37 &0.0078 &0.1350 &0.1262 \\
0.38 &0.0049 &0.1249 &0.1123 \\
0.39 &0.0030 &0.1188 &0.1028 \\
0.40 &0.0018 &0.1127 &0.0933 \\ \hline
\end{tabular}
\end{center}
\begin{center}
\caption{Fermi momentum distribution of proton inside $^{3}\mbox{He}$ according to three different models: analytic formula (equation \ref{eq:4.1}) (left), AV18 NN model+UrbanaIX three nucleon interaction (middle) and CDB2000 NN model+Tucson-Melbourne three nucleon interaction (right). The distributions were normalized to unity in the momentum range from 0 to 0.4 GeV/c.}
\end{center}
\end{table}

\newpage
\begin{table}[h!]
\footnotesize
\begin{center}
\begin{tabular}{|c|c|c|c|}\hline
$p_{F}$ &\multicolumn{3}{|c|}{$f(p_F)$ [c/GeV]}\\
\cline{2-4} 
[Gev/c] &anal. formula~\cite{Abdullin} &AV18+UrbIX~\cite{Nogga1,Nogga2} &CDB2000+TM~\cite{Nogga1,Nogga2} \\ 
\hline
0.01 &0.4788 &0.5316 &0.4392 \\
0.02 &1.7861 &1.9188 &1.6039 \\
0.03 &3.5812 &3.7094 &3.1551 \\
0.04 &5.4299 &5.4606 &4.7375 \\
0.05 &6.9442 &6.7807 &6.0219 \\
0.06 &7.8847 &7.5975 &6.9043 \\
0.07 &8.1972 &7.9148 &7.3554 \\
0.08 &7.9806 &7.8206 &7.4244 \\
0.09 &7.4157 &7.4358 &7.2021 \\
0.10 &6.6911 &6.8712 &6.7825 \\
0.11 &5.9512 &6.2074 &6.2400 \\
0.12 &5.2771 &5.5055 &5.6248 \\
0.13 &4.6944 &4.8276 &5.0082 \\
0.14 &4.1926 &4.1805 &4.4049 \\
0.15 &3.7476 &3.5912 &3.8388 \\
0.16 &3.3361 &3.0738 &3.3299 \\
0.17 &2.9434 &2.5893 &2.8478 \\
0.18 &2.5639 &2.2059 &2.4546 \\
0.19 &2.1996 &1.8456 &2.0831 \\
0.20 &1.8557 &1.5583 &1.7797 \\
0.21 &1.5384 &1.2975 &1.5023 \\
0.22 &1.2531 &1.0913 &1.2781 \\
0.23 &1.0029 &0.9036 &1.1073 \\
0.24 &0.7886 &0.7620 &0.9136 \\
0.25 &0.6096 &0.6261 &0.7606 \\
0.26 &0.4633 &0.5332 &0.6523 \\
0.27 &0.3462 &0.4403 &0.5440 \\
0.28 &0.2545 &0.3758 &0.4662 \\
0.29 &0.1840 &0.3176 &0.3951 \\
0.30 &0.1309 &0.2684 &0.3341 \\
0.31 &0.0916 &0.2338 &0.2892 \\
0.32 &0.0631 &0.1991 &0.2443 \\
0.33 &0.0428 &0.1769 &0.2137 \\
0.34 &0.0286 &0.1574 &0.1862 \\
0.35 &0.0188 &0.1391 &0.1601 \\
0.36 &0.0122 &0.1287 &0.1438 \\
0.37 &0.0078 &0.1183 &0.1274 \\
0.38 &0.0049 &0.1095 &0.1131 \\
0.39 &0.0030 &0.1042 &0.1034 \\
0.40 &0.0018 &0.0989 &0.0937 \\ \hline
\end{tabular}
\end{center}
\begin{center}
\caption{Fermi momentum distribution of neutron inside T according to three different models:  analytic formula (equation \ref{eq:4.1}) (left), AV18 NN model+UrbanaIX three nucleon interaction (middle) and CDB2000 NN model+Tucson-Melbourne three nucleon interaction (right). The distributions were normalized to unity in the momentum range from 0 to 0.4 GeV/c.}
\end{center}
\end{table}

\newpage
\begin{table}[h!]
\scriptsize 
\begin{center}
\begin{tabular}{|c|c|c|c|}\hline
$p_{F}$ &\multicolumn{3}{|c|}{$f(p_F)$ [c/GeV]}\\
\cline{2-4} 
[Gev/c] &anal. formula~\cite{Hejny} &AV18+UrbIX~\cite{Nogga2} &CDB2000+TM~\cite{Nogga2} \\
\hline
0.01  &0.0465  &0.1456	&0.1429 \\
0.02  &0.1839  &0.5488	&0.5345 \\
0.03  &0.4065  &1.1655	&1.1395 \\
0.04  &0.7053  &1.9169	&1.8881 \\
0.05  &1.0679  &2.7746	&2.7056 \\
0.06  &1.4798  &3.5499	&3.5139 \\
0.07  &1.9247  &4.2831	&4.2348 \\
0.08  &2.3855  &4.9130  &4.8657 \\
0.09  &2.8450  &5.3469	&5.3272 \\
0.10  &3.2868  &5.6795	&5.6216 \\
0.11  &3.6956  &5.7905	&5.7969 \\
0.12  &4.0585  &5.7862	&5.7750 \\
0.13  &4.3646  &5.6802  &5.6728 \\
0.14  &4.6062  &5.4062	&5.4463 \\
0.15  &4.7782  &5.1322	&5.1420 \\
0.16  &4.8784  &4.7547	&4.8117 \\
0.17  &4.9075  &4.3501	&4.4151 \\
0.18  &4.8685  &3.9455	&4.0184 \\
0.19  &4.7666  &3.5502	&3.6288 \\
0.20  &4.6087  &3.1561	&3.2422 \\
0.21  &4.4029  &2.7620	&2.8557 \\
0.22  &4.1581  &2.4289	&2.5326 \\
0.23  &3.8835  &2.1320  &2.2222 \\
0.24  &3.5881  &1.8356	&1.9110 \\
0.25  &3.2807  &1.5385	&1.6683 \\
0.26  &2.9692  &1.3591	&1.4551 \\
0.27  &2.6607  &1.1797	&1.2418 \\
0.28  &2.3611  &1.0003	&1.0518 \\
0.29  &2.0754  &0.8093	&0.9250 \\
0.30  &1.8073  &0.7337	&0.7983 \\
0.31  &1.5593  &0.6469	&0.6716 \\
0.32  &1.3332  &0.5602	&0.5779 \\
0.33  &1.1298  &0.4734	&0.5121 \\
0.34  &0.9490  &0.4194	&0.4463 \\
0.35  &0.7902  &0.3868	&0.3805 \\
0.36  &0.6523  &0.3542	&0.3332 \\
0.37  &0.5339  &0.3216	&0.3026 \\
0.38  &0.4333  &0.2889	&0.2720 \\
0.39  &0.3488  &0.2759	&0.2415 \\
0.40  &0.2784  &0.2665	&0.2149 \\ 
0.41  &0.2204  &0.2570	&0.2012 \\
0.42  &0.1730  &0.2475	&0.1875 \\
0.43  &0.1348  &0.2380	&0.1737 \\
0.44  &0.1041  &0.2331	&0.1601 \\
0.45  &0.0798  &0.2295	&0.1510 \\
0.46  &0.0607  &0.2260	&0.1439 \\
0.47  &0.0458  &0.2224	&0.1368 \\
0.48  &0.0343  &0.2189	&0.1297 \\
0.49  &0.0254  &0.2153	&0.1226 \\
0.50  &0.0187  &0.2133	&0.1177 \\ \hline
\end{tabular}
\end{center}
\begin{center}
\caption{Fermi momentum distribution of neutron inside $^{4}\mbox{He}$ according to three different models: analytic formula (equation \ref{eq:4.2}) (left), AV18 NN model+UrbanaIX three nucleon interaction (middle) and CDB2000 NN model+Tucson-Melbourne three nucleon interaction (right). The distributions were normalized to unity in the momentum range from 0 to 0.5 GeV/c.}
\end{center}
\end{table}

  
\newpage
\thispagestyle{plain}
  \section{Acceptance-supplement}

\vspace{0.5cm}

\noindent
The following tables show all possible combinations of particle registration for each of considered reactions. Abbreviations 'FD', 'CD-MDC' and 'CD-EC' denote Forward Detector, Mini Drift Chamber and Electromagnetic Calorimeter of WASA-at-COSY detector, respectively. 
  
\noindent
\begin{table}[h!]
\begin{footnotesize}
\begin{center}
\begin{tabular}{|c|c|c|c|c|}\hline
particle  &case 1 &case 2 &case 3 &case 4\\
\hline 
$^{3}\hspace{-0.03cm}\mbox{He}$  &FD &FD &FD &FD \\
proton  &FD &CD-MDC &FD &CD-MDC \\
$\pi^{-}$ &FD &CD-MDC &CD-MDC &FD \\
\hline
\end{tabular}
\end{center}
\begin{center}
\caption{Possibilities of particles registration in the $dd\rightarrow(^{4}\hspace{-0.03cm}\mbox{He}$-$\eta)_{bs}\rightarrow$ $^{3}\hspace{-0.03cm}\mbox{He} p \pi{}^{-}$ reaction.}
\end{center}
\end{footnotesize}
\end{table}

\vspace{-0.5cm} 

\noindent
\begin{table}[h!]
\begin{footnotesize}
\begin{center}
\begin{tabular}{|c|c|c|c|c|c|}\hline
particle  &case 1 &case 2 &case 3 &case 4 &case 5        \\
\hline                 
deuteron  &FD &CD-MDC  &FD      &FD        &FD             \\
proton1   &FD &CD-MDC  &CD-MDC  &CD-MDC    &CD-MDC          \\
proton2   &FD &CD-MDC  &CD-MDC  &FD        &CD-MDC        \\
$\pi^{-}$ &FD &CD-MDC  &CD-MDC  &FD        &FD             \\
\hline
\end{tabular}
\end{center}
\begin{center}
\begin{tabular}{|c|c|c|c|c|c|}\hline
case 6 &case 7 &case 8 &case 9 &case 10 &case 11 \\
\hline                 
FD       &FD      &FD  &FD     &CD-MDC   &CD-MDC    \\
CD-MDC   &FD      &FD   &FD     &CD-MDC   &FD    \\
FD       &CD-MDC  &FD   &CD-MDC &CD-MDC   &CD-MDC   \\
CD-MDC   &CD-MDC  &CD-MDC &FD     &FD       &CD-MDC  \\
\hline
\end{tabular}
\end{center}
\begin{center}
\begin{tabular}{|c|c|c|c|c|}\hline
case 12 &case 13 &case 14 &case 15 &case 16\\
\hline     
CD-MDC   &CD-MDC  &CD-MDC   &CD-MDC   &CD-MDC \\
CD-MDC   &FD      &FD       &FD       &CD-MDC \\
FD       &FD      &CD-MDC   &FD       &FD   \\
CD-MDC   &CD-MDC  &FD       &FD       &FD   \\
\hline
\end{tabular}
\end{center}
\begin{center}
\caption{Possibilities of particles registration in the $dd\rightarrow(^{4}\hspace{-0.03cm}\mbox{He}$-$\eta)_{bs}\rightarrow$ $d p p \pi{}^{-}$ reaction.}
\end{center}
\end{footnotesize}
\end{table}

\newpage    
\noindent
\begin{table}[h!]
\begin{footnotesize}
\begin{center}
\begin{tabular}{|c|c|c|}\hline
particle  &case 1 &case 2 \\
\hline 
deuteron  &FD      &FD  \\
proton    &CD-MDC  &FD  \\
gamma1    &CD-EC   &CD-EC  \\
gamma2    &CD-EC   &CD-EC   \\
\hline
\end{tabular}
\end{center}
\vspace{-0.8cm}
\begin{center}
\caption{Possibilities of particles registration in the $pd\rightarrow(^{3}\hspace{-0.03cm}\mbox{He}$-$\eta)_{bs}\rightarrow$ $d p \gamma \gamma$ reaction.}
\end{center}
\end{footnotesize}
\end{table}

\vspace{-0.5cm} 

\noindent
\begin{table}[h!]
\begin{footnotesize}
\begin{center}
\begin{tabular}{|c|c|c|c|c|c|}\hline
particle  &case 1 &case 2 &case 3 &case 4 &case 5        \\
\hline                 
proton1  &FD &CD-MDC  &FD      &FD        &FD             \\
proton2   &FD &CD-MDC  &CD-MDC  &CD-MDC    &CD-MDC          \\
proton3   &FD &CD-MDC  &CD-MDC  &FD        &CD-MDC        \\
$\pi^{-}$ &FD &CD-MDC  &CD-MDC  &FD        &FD             \\
\hline
\end{tabular}
\end{center}
\begin{center}
\begin{tabular}{|c|c|c|c|c|c|}\hline
case 6 &case 7 &case 8 &case 9 &case 10 &case 11 \\
\hline                 
FD       &FD      &FD  &FD     &CD-MDC   &CD-MDC    \\
CD-MDC   &FD      &FD   &FD     &CD-MDC   &FD    \\
FD       &CD-MDC  &FD   &CD-MDC &CD-MDC   &CD-MDC   \\
CD-MDC   &CD-MDC  &CD-MDC &FD     &FD       &CD-MDC  \\
\hline
\end{tabular}
\end{center}
\begin{center}
\begin{tabular}{|c|c|c|c|c|}\hline
case 12 &case 13 &case 14 &case 15 &case 16\\
\hline     
CD-MDC   &CD-MDC  &CD-MDC   &CD-MDC   &CD-MDC \\
CD-MDC   &FD      &FD       &FD       &CD-MDC \\
FD       &FD      &CD-MDC   &FD       &FD   \\
CD-MDC   &CD-MDC  &FD       &FD       &FD   \\
\hline
\end{tabular}
\end{center}
\vspace{-0.8cm}
\begin{center}
\caption{Possibilities of particles registration in the $pd\rightarrow(^{3}\hspace{-0.03cm}\mbox{He}$-$\eta)_{bs}\rightarrow$ $p p p \pi{}^{-}$ reaction.}
\end{center}
\end{footnotesize}
\end{table}
  
\vspace{-0.5cm}  
  
\noindent
\begin{table}[h!]
\begin{footnotesize}
\begin{center}
\begin{tabular}{|c|c|}\hline
particle  &case 1  \\
\hline 
proton$_{sp}$  &START and STOP DETECTORS	  \\
deuteron    &START and STOP DETECTORS  \\
proton     &START and STOP DETECTORS  \\
$\pi^{-}$    &START and STOP DETECTORS   \\
\hline
\end{tabular}
\end{center}
\vspace{-0.8cm}
\begin{center}
\caption{Possibilities of particles registration in the $dd\rightarrow p_{sp}(\mbox{T}$-$\eta)_{bs}\rightarrow$ $ p_{sp}d p \pi{}^{-}$ reaction.}
\end{center}
\end{footnotesize}
\end{table} 
  

\newpage
\pagestyle{fancy}
\fancyhf{} 
\fancyhead[LE,RO]{\textbf{\thepage}}
\fancyhead[RE]{\small\textbf{{References}}} 

\newpage
\thispagestyle{plain}

\end{document}